\documentclass[12pt]{article}
\pdfoutput=1
\usepackage{color}
\def\pier #1{{\color{red} #1}}
\def\betti #1{{\color{blue} #1}}
\def\ale #1{{\color{magenta} #1}}
\def\pier#1{{#1}}
\def\betti#1{{#1}}
\def\ale#1{{#1}}

\def\revis #1{{\color{red} #1}}
\def\revis #1{{#1}}

\def\last #1{{\color{red} #1}}
\def\last #1{{#1}}

\usepackage{amssymb}
\usepackage{amsfonts}
\usepackage{cite}
\usepackage{graphicx}

\usepackage{tabularx}
\newtheorem{theorem}{Theorem}[section]

\newtheorem{proposition}[theorem]{Proposition}

\input{tcilatex}
\begin{document}

\renewcommand{\theequation}{\arabic{section}.\arabic{equation}}

\begin{center}
\pier{\Large Chemotaxis-inspired PDE model for airborne infectious 
\\[1mm] disease
transmission: analysis and simulations}

\bigskip

Pierluigi Colli\footnote{%
Dipartimento di Matematica \textquotedblleft F. Casorati\textquotedblright ,
Universit\`{a} di Pavia and Research Associate at the IMATI -- C.N.R. Pavia,
via Ferrata 5, 27100 Pavia, Italy
}

E-mail: pierluigi.colli@unipv.it

\medskip

Gabriela Marinoschi\footnote{%
Gheorghe Mihoc-Caius Iacob Institute of Mathematical Statistics and Applied
Mathematics of the Romanian Academy, Calea 13 Septembrie 13, 050711
Bucharest, Romania}

E-mail: gabriela.marinoschi@acad.ro

\medskip

Elisabetta Rocca$^{1}$

E-mail: elisabetta.rocca@unipv.it

\medskip

Alex Viguerie\footnote{%
Gran Sasso Science Institute,
Viale Francesco Crispi 27,
L'Aquila, AQ 67100, Italy}

E-mail: alexander.viguerie@gssi.it
\end{center}

\noindent 
{\bf Abstract.} Partial differential equation (PDE) models for infectious disease have received renewed interest in recent years. Most models of this type extend classical compartmental formulations with additional terms accounting for spatial dynamics, with Fickian diffusion being the most common such term. However, while diffusion may be appropriate for modeling vector-borne diseases, or diseases among plants or wildlife, the spatial propagation of airborne diseases in human populations is heavily dependent on human contact and mobility patterns, which are not necessarily well-described by diffusion. By including an additional chemotaxis-inspired term, in which the infection is propagated along the positive gradient of the susceptible population (from regions of low- to high-density of susceptibles), one may provide a more suitable description of these dynamics. This article introduces and analyzes a mathematical model of infectious disease incorporating a modified chemotaxis-type term. The model is analyzed mathematically and \betti{the} well-posedness \betti{of the resulting PDE system} is demonstrated. A series of numerical simulations are provided, demonstrating the ability of the model to naturally capture important phenomena not easily observed in standard diffusion models, including \ale{propagation over long spatial distances over short time scales} and the emergence of localized infection hotspots.
\medskip

\noindent \pier{\bf Keywords:} nonlinear parabolic equations, reaction-diffusion systems, chemotaxis, existence and uniqueness of solutions, epidemic models, COVID-19.
\medskip

\pier{\noindent {\bf AMS (MOS) Subject Classification:} 
35K55, 
35K57, 
35Q92, 
46N60, 
92C17, 
92D30.} 

\section{Introduction}

\setcounter{equation}{0}

Mathematical models have long found application in the modeling and study of infectious disease, dating back to Daniel Bernoulli's model of smallpox vaccination in the 1700s \cite{colombo2015smallpox}. Over the following centuries, mathematical models have been found numerous applications in in epidemiology and public health. The most commonly employed models are \textit{compartmental models}, based on systems of ordinary differential equations (ODEs), following the framework of the susceptible-infectious-recovered (hereafter abbreviated as \textit{SIR}) model, itself originating from a
 special case of the more general model introduced by Kermack and
McKendrick \cite{kermack1927contribution} (see \cite{breda2012formulation}
for a more modern treatment, as well as\pier{\cite{GM-AMO, GM-DCDS}}). These models divide the population into different compartments based on disease stage.
Such models are easily implemented and analyzed, and have relatively
low computational overhead. However, they do not naturally incorporate
spatial information.

To overcome this limitation, analogous models based on partial differential
equations (PDEs) have been proposed \cite{murray2003mathematical,
schiesser2018mathematical}. These models have historically found success,
particularly in modeling vector-borne diseases and diseases spread among
wildlife \cite{murray2003mathematical, keller2013numerical}. These models are typically of reaction-diffusion type, and model the spatial propagation of disease as a Fickian diffusive process, in which the infection moves along a negative gradient. Reaction-diffusion models have also been used to model infectious diseases
in human populations. \last{While much recent work in this area has been applied
towards COVID-19 \cite{ grave2022modeling,
bertaglia2021hyperbolic, bertaglia2024newtrends, burini2024epidemics,
auricchio2022well, CGMR-1, CGM}}%
, PDE models have also been used for other diseases, or described in more
general settings \cite{li2009modeling, germann2006mitigation,
albi2022kinetic, ramaswamy2021comprehensive}. \revis{We also note that distinct approaches using PDEs to model infectious disease at smaller-scale, have incorporated principles from the kinetic modeling of crowd dynamics \cite{kim2021kinetic, agnelli2023spatial}.}

While PDE models offer the advantage of describing spatial dynamics, their
use in modeling infectious disease in human populations remains limited, for
reasons both practical and epidemiological. From the practical point of
view, PDEs require more computational resources, more data, and more effort to implement and solve compared to their ODE
counterparts. Additionally, their mathematical analysis is often
complicated, and the definition of important epidemiological quantities
(such as the reproduction number) is not straightforward \cite%
{auricchio2022well, albi2022kinetic}.

More seriously, however, there are the issues from the epidemiological point
of view: it is not clear that a diffusive process provides a suitable
description of infectious disease in a human population, given the complex,
multiscale nature of human mobility \cite{findlater2018human, belik2011natural}. Diffusion models do not provide a natural
description of nonlocal transmission dynamics. In contrast, ODE models are
quite flexible in this regard, and one may model (potentially distant)
geographic regions via an additional compartmental stratification. Thus, in
addition to their computational and mathematical attractiveness, carefully
designed ODE models may also provide more suitable spatial description
compared to a PDE model in certain settings. This is particularly well-suited for models of sexually transmitted diseases, where transmission does not depend strongly on local population density \cite{tatapudi2022evaluating, blackwood2018introduction}. Such models are, however, very data-hungry and their \pier{parametrization} can become challenging as the spatial resolution increases. This causes difficulties when applying this approach to airborne infectious diseases in human populations, as transmission may depend heavily on local population density patterns \cite{sy2021population, hu2013scaling}. 

The inability of reaction-diffusion models of infectious disease to
incorporate nonlocal mobility is a well-known limitation, and significant
effort has been applied towards resolving it. Models incorporating
bilaplacian \cite{murray2003mathematical} and fractional diffusion \cite%
{zhao2023spatiotemporal} terms may naturally reproduce nonlocal behavior; however, both
their implementation and analysis \betti{are} different. Reaction-diffusion models
have also been extended by, for example, incorporating additional advection
terms (possibly endowed with a network structure \cite%
{bertaglia2021hyperbolic, albi2022kinetic, ramaswamy2021comprehensive}) or
mass-transfer operators \cite{grave2022modeling} to allow for nonlocal
effects.

While such efforts have produced promising results, the focus in these
instances is to better reproduce the \textit{nonlocal} behavior. However,
another equally important question, whether diffusion is in fact an
appropriate description of the \textit{local} spatial dynamics of an
airborne infectious disease in human populations, has not received similar
attention. Indeed, while additional care is occasionally given in defining
the diffusive process (for instance, preferential or nonlinear diffusion 
\cite{grave2022modeling, viguerie2022coupled}),
such considerations affect only the \textit{rate} and \textit{areas} in
which diffusion occurs. The underlying idea that an infectious disease in a
human population should \textit{diffuse}, that is, travel from regions of
high-infection density to regions of low-infection density, as the limit of
a Brownian motion, is discussed in few works (see e.g., \cite{fitz-18,
fitz-21, auricchio2022well, CGMR-1, CGM}).

In the present work, we consider that diffusion-based models of airborne
infectious diseases in humans are also limited in their description of 
\textit{local} dynamics, and that such limitations are of similar, if not greater, importance as their well-known problems with describing \textit{nonlocal} dynamics. In particular, the local transmission processes modeled by reaction-diffusion PDE models have difficulty in producing some or all of the following phenomena:
\begin{enumerate}
        \item rapid transmission over large geographic distances over short time-scales;
        \item variable transmission rates and spatial propagation arising due to differences in local population density patterns;
        \item the rapid appearance of the disease, and subsequent sustained transmission, in major urban centers, even in the absence of an infected population at initialization.
\end{enumerate}
While these phenomena are observed in airborne infectious diseases in human populations, standard reaction-diffusion PDE models generally require time- and/or space-dependent parameterizations in order to reproduce them.

In order to overcome these limitations, we introduce a
new model to better-capture the spatial propagation observed in real-world
human contagion. Specifically, the model does not only consider movement of
the disease along the negative gradient of infection density, but also the
nonlinear movement along the \textit{positive gradient} of susceptible
density; that is, from regions of low-to-high susceptible concentration.
While the underlying physical process is not the same, we note that this
model bears mathematical similarity to well-known models of \textit{
chemotaxis}, a process in which a biological species is attracted or
repelled in response to a chemical stimulus \cite{gajewski1998global,
bellomo2022chemotaxis, marinoschi2013well, GM-NA, negreanu2014two,
chalub2004kinetic}.

The article is outlined as follows. We first introduce the mathematical
model, characterizing transmission, diffusion and a \textit{%
chemotactic-inspired movement}, as previously explained\pier{. We} describe its basic
properties and explain the underlying intuition. After establishing the
necessary formality and notation, we proceed to \pier{examine} the mathematical
well-posedness of the model. Following this analysis, we prove the existence
of a solution (\pier{see} Theorem~\ref{t-due}) \pier{with} its properties, as e.g., positiveness.
Then, after \pier{showing} an additional regularity of the solution in
Proposition~\ref{prop-quattro}, we prove also its uniqueness in Theorem~\ref%
{t-cinque} \pier{in the case when} the diffusion coefficient for susceptibles is constant. \pier{Moreover, in Theorem~\ref{t-sei} we investigate the asymptotic behavior of our system as the diffusion coefficients for susceptible and removed compartments tend to $0$ by proving the convergence to the resulting mixed ODE-PDE system.}
Then,
we briefly discuss some important considerations regarding the numerical
solution of the model.  We then perform a proof-of-concept simulation and show that the model  provides better qualitative agreement with surveillance data compared to a purely diffusive model. Additionally, we show that the chemotaxis model recreates important dynamics not seen in the pure diffusion model, \ale{including transmission over long geographic distances over short time scales.} Furthermore, we show that these dynamics will occur naturally, and do not require spatiotemporal variation of model parameters. We conclude by summarizing the presented results and suggesting several directions for future work in this area.


\section{Mathematical model}

\setcounter{equation}{0}

Let us begin in a general setting, by \betti{taking a bounded connected} domain $%
\Omega \subset \pier{\mathbb{R}^{d}}$, $d=1,2,3,$ and let $i(x,t)$ describe the density of
infected individuals at a point $x$ in $\Omega $ at a time $t$. Next, we
consider a \betti{bounded connected}, arbitrary control region $\Omega _{a}\subset
\Omega $, with $\partial \Omega _{a}$ sufficiently smooth. The overall
concentration $I_{a}(t)$ of infection in $\Omega _{a}$ is given~by 
\begin{equation}
I_{a}(t)=\int_{\Omega _{a}}i(x,t)\,dx.  \label{Ia}
\end{equation}%
Following the standard reasoning from conservation laws, $I_{a}(x,t)$ can
change only due~to:

\begin{enumerate}
\item[(A)] Internal sources and sinks, or

\item[(B)] Infected population entering (or exiting) $\Omega_a$ from across the boundary $%
\partial \Omega_a$.
\end{enumerate}

Physically, (A) above refers to the generation of new infections from inside 
$\Omega _{a}$, also known as \textit{incidence} and denoted as $\lambda
(x,t) $, and removals from the infected population, either from recovery
from infection or death, within $\Omega _{a}$. We assume, for simplicity,
that all removals are recoveries and occur at a rate $\phi $.

Regarding (B), we may denote the movement by infected individuals across the
boundaries as a flux $\mathbf{j}$, assumed generic for the moment. From
standard arguments, then we have that 
\begin{equation}
\partial _{t}\int_{\Omega _{a}}i(x,t)\,dx=\int_{\Omega _{a}}\left( \lambda
(x,t)-\phi i(x,t)\right) \,dx-\int_{\partial \Omega _{a}}\mathbf{j}\cdot 
\mathbf{n}_{a}\,d\sigma ,  \label{continuityEqn}
\end{equation}%
where $\mathbf{n}_{a}$ denotes the outward normal unit vector to $\partial
\Omega _{a}$. Applying the divergence theorem, moving the time derivative
inside the integral, and rearranging the latter equality implies 
\begin{equation}
\int_{\Omega _{a}}\left( \betti{\partial _{t} i}(x,t)-\lambda
(x,t)+\phi i(x,t)+\nabla \cdot \mathbf{j}(x,t)\right) \,dx=0.  \label{div}
\end{equation}%
Since this must hold for any choice of $\Omega _{a}$, we thus have 
\begin{equation}
\partial _{t}i=\lambda -\phi i-\nabla \cdot \mathbf{j}  \label{intro1}
\end{equation}%
over the whole of $\Omega $. We must specify how we define $\lambda $ and $%
\mathbf{j}$, with the latter being the primary focus of the present work.
For the former, we will assume a standard frequency-dependent formulation \cite{blackwood2018introduction} , given as: 
\begin{equation}
\lambda (x,t)=\pier{\beta (x,t)}\frac{ i}{n}s,  \label{lambda}
\end{equation}%
where $s(x,t)$ refers to the susceptible population at $(x,t)$, $\beta (x,t)$
is the contact rate (units 1/time), and $n(x,t)=s(x,t)+i(x,t)+r(x,t)$ is the
total living population at $(x,t)$. Here, $r(x,t)$ represents the recovered
population. We briefly remark that other choices of $\lambda $ may be more
suitable for certain applications, including, among others,
density-dependent or Holling-type formulations. We refer the
reader to \cite{blackwood2018introduction} for further discussion of these issues. We now focus our attention on the choice of $\mathbf{j}$.

\subsection{Standard flux choice: Fickian diffusion}

For $\mathbf{j}$, the most common choice is to assume the \revis{\textit{Fickian
diffusion}} flux -- that is, the flux is proportional to the negative
gradient of infected density. Put simply, this assumption states that the
infected population will tend towards a uniform concentration, moving from
regions of high-infection density to low-infection density. Mathematically,
we may describe this as 
\begin{equation}
\mathbf{j}=-\nu _{i}\nabla i,  \label{Fick}
\end{equation}%
with $\nu _{i}(x,t)$ being the \textit{diffusion coefficient}.

{Another way to interpret this term is as a stochastic process. Letting $\textbf{X}_t$ denote the position of a member $\widetilde{i}$ of the population $i$ in space, then one may describe the evolution of $\textbf{X}_t$ in time as:    }
\begin{equation}\label{stochProc}
d\textbf{X}_t = \sqrt{2\nu_i}d\textbf{W}_t,
\end{equation}
{where $\textbf{W}_t$ denotes the standard Wiener process. The \textit{Fokker-Planck} equation corresponding to (\ref{stochProc}) then reads \cite{gardiner2009Stochastic, SSalsa}:}
\begin{equation}
\partial_t p(x,t) = \nabla \cdot \lbrack \nu_i \nabla p(x,t)  \rbrack,
\end{equation}
{where $p(x,t)$ is the probability density function such that, for a subset $\Omega_a \subseteq \Omega$:}
\begin{equation}\label{positionalDensity}
P(\widetilde{i} \in \Omega_a | t=t_a ) = \int_{\Omega_a} p(x,t_a) dx.
\end{equation}
{By extending this reasoning over the entire population $i$, one arrives at the expression for the flux (\ref{Fick}). In this way, we can regard the PDE models as aggregated continuum limits of agent-based approaches. Indeed, such reasoning can be applied to a wide range of mathematical models for infectious disease \cite{zanella2024derivation}.}

Following this reasoning for additional compartments, and \pier{taking advantage of} a standard 
\textit{frequency-dependent} formulation of $\lambda $, one may derive the \textit{susceptible-infected-removed~(SIR)} \pier{system for}
susceptible, infected, removed population densities $s(x,t)$, $i(x,t)$, $r(x,t)$\pier{, respectively:}  
\begin{equation}
\partial _{t}s=-\frac{\beta i}{s+i+r}s+\nabla \cdot \left( \nu _{s}\nabla
s\right) ,\text{ in }Q:=\Omega \times (0,\,T),  \label{I1}
\end{equation}%
\begin{equation}
\partial _{t}i=\frac{\beta i}{s+i+r}s-\phi i+\nabla \cdot \left( \nu
_{i}\nabla i\right) ,\text{ in }Q,  \label{I2}
\end{equation}%
\begin{equation}
\partial _{t}r=\phi i+\nabla \cdot \left( \nu _{r}\nabla r\right) ,\text{ in 
}Q,  \label{I3}
\end{equation}%
\begin{equation}
\nabla s\cdot \mathbf{n}\mathbf{=}\nabla i\cdot \mathbf{n=}\nabla r\cdot 
\mathbf{n}=0,\text{ on }\Sigma :=\partial \Omega \,\times (0,T),  \label{I4}
\end{equation}%
\begin{equation}
s(0)=s_{0},\,i(0)=i_{0},\,r(0)=r_{0},\text{ in }\Omega \pier{.}  \label{I5}
\end{equation}%
\pier{We recall that $\Omega $ denotes} a bounded \betti{domain} of $\mathbb{R}^{d}$, $d=1,2,3$,
with a smooth boundary $\partial \Omega $\pier{. Here, $\mathbf{n}$ stands for} the outward
normal unit vector to the boundary $\partial \Omega $ and $T>0$ represents a
final time.

\subsection{Flux definition 2: Chemotaxis-inspired}

Let us consider an alternative choice for the flux. We assume that the
spatial evolution of the infectious disease depends primarily on two
processes:

\begin{enumerate}
\item An undirected component, modeled as a random walk. As mentioned the
previous example, in the continuous limit, this is described by a diffusion
operator.

\item A \textit{directed} component, which corresponds to mobility which is
predictable, and not random.
\end{enumerate}

To model this second process, we assume the following:

\begin{itemize}
\item The spatial propagation of an infectious disease depends on human
mobility, in particular how this mobility relates to \textit{contacts}. In
general, contacts will increase moving from regions of lower to higher
population density.

\item Higher levels of infection should, in general, serve to \textit{%
increase}, not decrease, the spatial propagation of the disease \textit{%
provided sufficient availability of susceptible individuals}.
\end{itemize}

Combining the two points, we therefore postulate that \textit{an airborne
infectious disease should propagate according to the density of susceptible
individuals}, and in particular \textit{will move from regions with a lower
concentration of susceptible individuals to
\pier{regions with} a higher concentration of
susceptible individuals}\pier{. The rate of this movement should be} also governed in
part by the concentration of infected individuals in the area.

Putting these pieces together suggests the following, alternative definition
for the flux: 
\begin{equation}
\mathbf{j}=-\nu _{i}\nabla i+\chi (i)\nabla s,  \label{altFlux}
\end{equation}%
where $\chi (i)$ is a strictly positive function of $i$. A possible \pier{specification for $\chi$}, {used in the current work,} is the \textit{Hill function given by:}
\begin{equation}
\pier{\chi (x,t,i)= \mu _{i}(x,t)} \dfrac{i}{1+i/C_{0}},  \label{Holling}
\end{equation}%
where $\mu _{i}>0$ is a chemotactic coefficient and $C_{0}>0$ is a \textit{%
capacity term} defining a level of population density at which saturation
occurs. Observe that when $i<<C_0$, $\chi(i)\approx \mu_i i$; hence, $\chi$ increases in proportion to $i$ for smaller $i$. However, as $i$ increases, this begins to change; note that for $i\approx C_0$, $\chi(i)$ will increase in proportion to $\mu_{i}i/2$\pier{.}  Finally, for $i>>C_0$ \pier{we see that}
\begin{equation}
\lim_{i\to \infty} \chi(i) = \lim_{i\to \infty}  \dfrac{\mu _{i}i}{1+i/C_{0}} = C_0 \mu_i .
\end{equation}
Hence, the parameter $C_0$ defines a level of infection density at which additional increases in infection density result in a reduced effect on the spatial dynamics, {due to the saturation effect of the Hill function}. Such terms may be explained physically by, for example, natural limitations on human mobility. 

The flux (\ref{altFlux}) suggests that the contact patterns of persons in a region should concentrate at local population maxima. As such, it can be seen as a space-continuous analogue of the discrete radiation model of population mobility \cite{s2012radiation}, which postulates that mobility between two localities depends on the populations of each region, as well as the population density in the intermediate areas between them.  This model was shown in \cite{d2024spatial} to provide an accurate description of COVID-19 spatial propagation in Italy.

As in the case of Fickian diffusion, (\ref{altFlux}) can also be understood in terms of stochastic processes, as the aggregated limit of a multiagent system. For simplicity, assume that $C_0$ is sufficiently large, such that:
$$  \chi(i) \approx \mu_i i. $$
Proceeding analogously as in (\ref{stochProc}), we consider the stochastic process:
\begin{equation}\label{stochProc2}
d\textbf{X}_t = -\mu_i \nabla s (\textbf{X}_t) + \sqrt{2 \nu_i } d\textbf{W}_t.
\end{equation}
The above expression differs from (\ref{stochProc}) due to the presence of the \textit{drift term} $-\mu_i\nabla s (\textbf{X}_t)$. This drift term governs the deterministic change in the average position, while the Wiener process introduces random fluctuations, whose magnitude determines the positional variance. From standard arguments, the Fokker-Planck equation corresponding to (\ref{stochProc2}) reads: 
\begin{equation}\label{fokkPlanck2}
\partial_t p(x,t) = \nabla  \cdot \left\lbrack  -\mu_i p(x,t) \nabla s + \nu_i \nabla p(x,t)\right\rbrack,
\end{equation}
{where $p(x,t)$ is as in (\ref{positionalDensity}). }

{Intuitively, the expression (\ref{fokkPlanck2}) states that, all else constant, the probability of finding infected individuals, over time, should increase as susceptible density increases. This makes sense, as regions with lower susceptible density are either low population-density regions or regions in which no susceptible persons remain to be infected. In either case, the probability of finding infected individuals in such locations is lower. In low population-density regions, overall transmission is reduced \cite{sy2021population, hu2013scaling}.  On the other hand, if a high-density area has a low concentration of susceptibles, this implies that the majority of the area's population has previously been infected. Therefore, most persons who frequent the area are likely to have already acquired infection. Extending this reasoning to the full population, one arrives at the flux 
term~(\ref{altFlux}).}

Proceeding as before, the \pier{general} modified model reads as follows:%
\begin{equation}
\partial _{t}s-\nabla \cdot (\nu _{s}\nabla s)+\frac{\beta i}{s+i+r}s=0,%
\text{ in }Q,  \label{eq-s}
\end{equation}%
\begin{equation}
\partial _{t}i-\nabla \cdot \left( \nu _{i}\nabla i-\chi (i)\nabla s\right)
+\phi i-\frac{\beta s}{s+i+r}i=0,\text{ in }Q,  \label{eq-i}
\end{equation}%
\begin{equation}
\partial _{t}r-\nabla \cdot (\nu _{r}\nabla r)-\phi i=0,\text{ in }Q,
\label{eq-r}
\end{equation}%
\begin{equation}
\nabla s\cdot \mathbf{n}=(\nu _{i}\nabla i-\chi (i)\nabla s)\cdot \mathbf{n}%
=\nabla r\cdot \mathbf{n}=0,\text{ on }\Sigma ,  \label{bc0}
\end{equation}%
\begin{equation}
s(0)=s_{0},\,i(0)=i_{0},\,r(0)=r_{0},\text{ in }\Omega .  \label{ci}
\end{equation}%
Looking at the boundary conditions (\ref{bc0}) we deduce in particular that 
\begin{equation}
\nabla i\cdot \mathbf{n}=0 \ \text{ on }\Sigma   \label{bc-i}
\end{equation}%
as well. \pier{From here on out we} will regard (\ref{I1})-(\ref{I5}) as simply a particular case of (\ref{eq-s})-(\ref{ci}) with $\mu_i=0$; note the 
capacity term $C_0$ no longer serves any purpose in such a case. A full list of the model parameters in 
(\ref{eq-s})-(\ref{ci}), their description, and their units \pier{are} provided in 
Table~\ref{tab:ModelParameters}. 

\begin{table}[tbp]
\begin{center}
\begin{tabular}{|c|c|c|}
\hline
Parameter & Name & Unit \\ \hline\hline
$\beta$ & Contact rate & Days$^{-1}$ \\ \hline
$\phi$ & Removal rate & Days$^{-1}$ \\ \hline
$\nu_j$ & Diffusion coefficient, compartment $j$ & km$^2$ $\cdot$ Days$^{-1}$ \\ \hline
$\mu_i$ & \pier{Chemotactic} coefficient & km$^2$ $\cdot$ Persons $\cdot$ Days$^{-1}$
\\ \hline
$C_0$ & Capacity/saturation term & Persons 
\\ \hline
\end{tabular}%
\end{center}
\caption{Description of model parameters for (\ref{eq-s})-(\ref{ci}) \pier{with the choice (\ref{Holling})}. Note that (\ref{I1})-(\ref{I5}) corresponds to the case of $\mu_i=0$. }\label{tab:ModelParameters}
\end{table}

\section{Mathematical analysis of the chemotaxis-inspired model}

\setcounter{equation}{0}

In this section, we present the mathematical results regarding the system (%
\ref{eq-s})-(\ref{ci}). In order to allow more generality to the system
parameters, which in principle can be different in space and vary in time we
assume that they are space and time dependent.

We begin by specifying some notation. Let 
\[
H=L^{2}(\Omega ),\text{ }V=H^{1}(\Omega ),\text{ }W=\{v\in H^{2}(\Omega );%
\text{ }\nabla v\cdot \mathbf{n=}0\text{ on }\partial \Omega \}. 
\]%
As usual, we identify the space $H$ \pier{with} its dual space $H^{\prime }$ and we
note that \betti{$W\subset V\subset H\equiv H^{\prime}\subset V^{\prime}$} with dense and
compact embeddings. We denote by $\left\langle \cdot ,\cdot \right\rangle $
the pairing between $V^{\prime }$ and $V.$

We assume the following hypotheses on the system parameters:

\begin{equation}
(\nu _{s},\nu _{i},\nu _{r})\in (L^{\infty }(Q))^{3},\text{ }0<\nu _{m}\leq
\nu _{s},\nu _{i},\nu _{r}\leq \nu _{M}\text{ a.e.~in }Q,\text{ }
\label{niu}
\end{equation}

\begin{equation}
\beta \in L^{\infty }(Q),\text{ }0\leq \beta \leq \beta _{M}\text{ a.e.~in }%
Q,\text{ }  \label{beta}
\end{equation}%
with $\nu _{m},$ $\nu _{M},$ $\beta _{M}$ constants,%
\begin{equation}
\phi \text{ is a positive coefficient,}  \label{fi}
\end{equation}%
\begin{equation}
\chi :Q\times \lbrack 0,+\infty )\rightarrow \mathbb{R}\text{ is a Carath%
\'{e}odory function, }  \label{chi1}
\end{equation}%
that is, $\chi (\cdot ,\cdot ,v)$ is measurable in $Q$ for all $v\geq 0$ and 
$\chi (x,t,\cdot )$ is continuous in $[0,+\infty )$ for a.e. $(x,t)\in Q.$

Moreover, there exist two positive constants $\chi _{M}$ and $\chi _{1,M}$
such that 
\begin{equation}
\left\vert \chi (x,t,v)\right\vert \leq \chi _{M},\text{ }\frac{\left\vert
\chi (x,t,v)\right\vert }{v}\leq \chi _{1,M}\text{ \ \ for all }v>0\text{
and a.e. }(x,t)\in Q.  \label{chi2}
\end{equation}%
An example of such a function may be \pier{(cf.~(\ref{Holling}))}
\[
\chi (x,t,v)=\chi _{0}(x,t)\frac{v}{v+c_{0}},\text{ with }\chi _{0}\in
L^{\infty }(Q),\text{ }\chi _{0}\geq 0\text{ a.e.~in }Q,\text{ }c_{0}>0. 
\]%
Finally, for the initial data we assume that 
\begin{equation}
(s_{0},i_{0},r_{0})\in (L^{2}(\Omega ))^{3},\text{ }0\leq s_{0}\leq s_{M},%
\text{ }i_{0}\geq 0,\text{ }r_{0}\geq 0\text{ a.e.~in }\Omega ,
\label{ci-cond}
\end{equation}%
with $s_{M}$ constant.

The aim is to prove that \pier{the initial-boundary value problem}~(\ref{eq-s})-(\ref{ci}) has a weak solution.
To this end\pier{,} we introduce an approximating system depending on a parameter $%
\varepsilon >0$, namely,
\begin{equation}
\partial _{t}s-\nabla \cdot (\nu _{s}\nabla s)+\frac{\beta i^{+}}{%
s^{+}+i^{+}+r^{+}+\varepsilon }\,s=0,\text{ in }Q,  \label{eq-s1}
\end{equation}%
\begin{equation}
\partial _{t}i-\nabla \cdot \left( \nu _{i}\nabla i-\chi (i^{+})\nabla
s\right) +\phi i-\frac{\beta s^{+}}{s^{+}+i^{+}+r^{+}+\varepsilon }\, i=0,%
\text{ in }Q,  \label{eq-i1}
\end{equation}%
\begin{equation}
\partial _{t}r-\nabla \cdot (\nu _{r}\nabla r)-\phi i=0,\text{ in }Q,
\label{eq-r1}
\end{equation}%
\begin{equation}
\nabla s\cdot \mathbf{n}=\nabla i\cdot \mathbf{n}=\nabla r\cdot \mathbf{n}=0,%
\text{ on }\Sigma ,  \label{bc1}
\end{equation}%
\begin{equation}
s(0)=s_{0},\,i(0)=i_{0},\,r(0)=r_{0},\text{ in }\Omega ,  \label{ci1}
\end{equation}%
where $v^{+}$ represents the positive part of $v=s,i,r.$ We note that the
last terms on the left-hand sides of the first two equations make sense now
because the denominators are positive. \pier{Similarly, the argument $i^+ $ of $\chi$ is nonnegative.}

First, we shall prove that system (\ref{eq-s1})-(\ref{ci1}) has a weak
solution $(s,i,r)$, with $s,i,r$ all nonnegative, so that $(s,i,r)$ solves
also the system 
\begin{equation}
\partial _{t}s-\nabla \cdot (\nu _{s}\nabla s)+\frac{\beta i}{%
s+i+r+\varepsilon }\,s=0,\text{ in }Q,  \label{eq-s-eps}
\end{equation}%
\begin{equation}
\partial _{t}i-\nabla \cdot \left( \nu _{i}\nabla i-\chi (i)\nabla s\right)
+\phi i-\frac{\beta s}{s+i+r+\varepsilon }\,i=0,\text{ in }Q,
\label{eq-i-eps}
\end{equation}%
\begin{equation}
\partial _{t}r-\nabla \cdot (\nu _{r}\nabla r)-\phi i=0,\text{ in }Q,
\label{eq-r-eps}
\end{equation}%
along with the boundary conditions (\ref{bc1}) and the initial conditions (%
\ref{ci1}).

Next, relying on appropriate {a} priori estimates we shall pass to the limit
as $\varepsilon \rightarrow 0$ in (\ref{eq-s-eps})-(\ref{eq-r-eps}) and prove
that the limit is a weak solution to the original system (\ref{eq-s})-(\ref%
{ci}). We note that the fraction $\frac{\beta is}{s+i+r}$ is well defined
whenever $s+i+r>0$ and it will be well specified when $s+i+r=0,$ as we shall
\pier{show} in the existence theorem.

We start with the auxiliary approximating system (\ref{eq-s1})-(\ref{ci1}).

\begin{theorem}
\label{t-uno} Under the assumptions (\ref{niu})-(\ref{ci-cond}) there exists
a triplet 
\begin{equation}
(s_{\varepsilon },i_{\varepsilon },r_{\varepsilon })\in (H^{1}(0,T;V^{\prime
})\cap C([0,T];H)\cap L^{2}(0,T;V))^{3}  \label{sol-eps}
\end{equation}%
\betti{satisfying the following notion of weak \pier{solution} to system (\ref{eq-s1})-(\ref{ci1}):
\[
\int_{0}^{T}\left\langle \partial _{t}s_{\varepsilon }(t),v(t)\right\rangle dt+\int_{Q}\nu
_{s}\nabla s_{\varepsilon }\cdot \nabla vdxdt+\int_{Q}\frac{\beta {i_{\varepsilon }}^{+}}{%
{s_{\varepsilon }}^{+}+{i_{\varepsilon }}^{+}+{r_{\varepsilon }}^{+}+\varepsilon }\,\pier{s_{\varepsilon }}vdxdt=0,
\]
\begin{eqnarray*}
&&\int_{0}^{T}\left\langle \partial _{t}i_{\varepsilon }(t),v(t)\right\rangle
dt+\int_{Q}\nu _{i}\nabla i_{\varepsilon }\cdot \nabla vdxdt+\int_{Q}\phi i_{\varepsilon }vdxdt  \nonumber
\\
&&-\int_{Q}\frac{\beta {s_{\varepsilon }}^{+}}{{s_{\varepsilon }}^{+}+\pier{i_{\varepsilon }}^{+}+%
{r_{\varepsilon }}^{+}+\varepsilon }\,\betti{i_{\varepsilon }}vdxdt=\int_{Q}\chi ({i_{\varepsilon }}%
^{+})\nabla s_{\varepsilon }\cdot \nabla vdxdt,  
\end{eqnarray*}
\[
\int_{0}^{T}\left\langle \partial _{t}r_{\varepsilon }(t),v(t)\right\rangle dt+\int_{Q}\nu
_{r}\nabla r_{\varepsilon }\cdot \nabla vdxdt-\int_{Q}\phi i_{\varepsilon }vdxdt=0,  
\]
for all $v\in L^2(0,T;V)$.}
Moreover, it
has the properties%
\begin{equation}
0\leq s_{\varepsilon }\leq s_{M},\text{ }i_{\varepsilon }\geq 0,\text{ }%
r_{\varepsilon }\geq 0\text{ a.e.~in }Q,  \label{sol-eps-pos}
\end{equation}%
whence it turns out that the solution $(s_{\varepsilon },i_{\varepsilon
},r_{\varepsilon })$ solves also \betti{a weak formulation of} equations (\ref{eq-s-eps})-(\ref{eq-r-eps})
with the boundary conditions (\ref{bc1}) and the initial conditions (\ref%
{ci1}), \betti{that is, for all $v\in L^2(0,T;V)$,
\[
\int_{0}^{T}\left\langle \partial_{t}s(t),v(t)\right\rangle dt+\int_{Q}\nu
_{s}\nabla s\cdot \nabla vdxdt+\int_{Q}\frac{\beta {i}}{{s}+{i}+{r}+\varepsilon}\,svdxdt=0,
\]
\begin{eqnarray*}
&&\int_{0}^{T}\left\langle \partial_{t}i(t),v(t)\right\rangle
dt+\int_{Q}\nu _{i}\nabla i\cdot \nabla vdxdt+\int_{Q}\phi ivdxdt  \nonumber
\\
&&-\int_{Q}\frac{\beta {s}}{{s}+{i}+{r}+\varepsilon }\,\betti{i}vdxdt=\int_{Q}\chi ({i})\nabla s\cdot \nabla vdxdt,  
\end{eqnarray*}
\[
\int_{0}^{T}\left\langle \partial _{t}r(t),v(t)\right\rangle dt+\int_{Q}\nu_{r}\nabla r\cdot \nabla vdxdt-\int_{Q}\phi ivdxdt=0. 
\]}%
\end{theorem}

\medskip

\noindent\textbf{Proof.} We apply the Schauder fixed point theorem, by
introducing the space 
\begin{equation}
X=(L^{2}(Q))^{3}  \label{X}
\end{equation}%
and the ball 
\begin{equation}
M=\{(s,i,r)\in X;\text{ }\left\Vert s\right\Vert _{L^{2}(Q)}+\left\Vert
i\right\Vert _{L^{2}(Q)}+\left\Vert r\right\Vert _{L^{2}(Q)}\leq R\},
\label{M}
\end{equation}%
with $R$ \pier{denoting} a large enough positive constant, which will be specified later.

We fix $(\overline{s},\overline{i},\overline{r})\in M$ and consider the
system%
\begin{equation}
\partial _{t}s-\nabla \cdot (\nu _{s}\nabla s)+\frac{\beta \overline{i}^{+}}{%
\overline{s}^{+}+\overline{i}^{+}+\overline{r}^{+}+\varepsilon }\,s=0,\text{
in }Q,  \label{eq-s2}
\end{equation}%
\begin{equation}
\partial _{t}i-\nabla \cdot \left( \nu _{i}\nabla i-\chi (\overline{i}%
^{+})\nabla s\right) +\phi i-\frac{\beta \overline{s}^{+}}{\overline{s}^{+}+%
\overline{i}^{+}+\overline{r}^{+}+\varepsilon }\,i=0,\text{ in }Q,
\label{eq-i2}
\end{equation}%
\begin{equation}
\partial _{t}r-\nabla \cdot (\nu _{r}\nabla r)-\phi i=0,\text{ in }Q,
\label{eq-r2}
\end{equation}%
with the boundary and initial conditions (\ref{bc1}), (\ref{ci1}) and prove
below that it has a solution $(s,i,r)$.

Then, we define the mapping 
\begin{equation}
\Phi :M\rightarrow X,\text{ }\Phi (\overline{s},\overline{i},\overline{r}%
):=(s,i,r)  \label{Fi}
\end{equation}%
and show that $\Phi (M)\subset M,$ $\Phi (M)$ is a compact set of $X$ and
that $\Phi $ is continuous.

Now we present the arguments that prove that $\Phi $ is well defined. Since,
due to (\ref{beta}), the coefficient of $s$ in (\ref{eq-s2}) satisfies 
\begin{equation}
0\leq \frac{\beta \overline{i}^{+}}{\overline{s}^{+}+\overline{i}^{+}+%
\overline{r}^{+}+\varepsilon }\leq \beta _{M}\text{ \ a.e.~in }Q,
\label{beta-bound}
\end{equation}%
it follows from the classical results concerning the well-posedness of
parabolic problems (see\pier{,} e.g., \cite{Lions}) that there exists a unique weak
solution%
\begin{equation}
s\in H^{1}(0,T;V^{\prime })\cap C([0,T];H)\cap L^{2}(0,T;V)  \label{s-reg}
\end{equation}%
to (\ref{eq-s2}), coupled with $\nabla s\cdot \mathbf{n}=0$ on $\Sigma $ and 
$s(0)=s_{0}.$

Moreover, by testing (\ref{eq-s2}) by $-s^{-}(t),$ where $s^{-}$ denotes the
negative part of $s,$ and integrating over $(0,t)$ we obtain%
\begin{eqnarray*}
&&\frac{1}{2}\left\Vert s^{-}(t)\right\Vert
_{H}^{2}+\int_{0}^{t}\!\!\int_{\Omega}\nu _{s}\left\vert \nabla
s^{-}\right\vert ^{2}dxd\tau  \\
&=&\frac{1}{2}\left\Vert s_{0}^{-}\right\Vert
_{H}^{2}-\int_{0}^{t}\!\!\int_{\Omega}\frac{\beta \overline{i}^{+}}{\overline{s}%
^{+}+\overline{i}^{+}+\overline{r}^{+}+\varepsilon }\left\vert
s^{-}\right\vert ^{2}dxd\tau .
\end{eqnarray*}%
Recalling that $s_{0}\geq 0$ a.e.~in~$\Omega $ and using (\ref{beta-bound})\pier{,}
we realize that the right-hand side is nonpositive and so we conclude that $%
s^{-}(t)=0$ for all $t\in \lbrack 0,T]$, whence $s\geq 0$ a.e.~in~$Q.$ We
proceed similarly by testing (\ref{eq-s2}) by the positive part $%
(s(t)-s_{M})^{+}$, integrating over $(0,t)$ and recalling again (\ref%
{ci-cond}) and (\ref{beta-bound}). Then, we obtain that 
\begin{eqnarray*}
&&\frac{1}{2}\left\Vert (s(t)-s_{M})^{+}\right\Vert
_{H}^{2}+\int_{0}^{t}\!\!\int_{\Omega}\nu _{s}\left\vert \nabla
(s-s_{M})^{+}\right\vert ^{2}dxd\tau  \\
&=&\frac{1}{2}\left\Vert (s_{0}-s_{M})^{+}\right\Vert
_{H}^{2}-\int_{0}^{t}\!\!\int_{\Omega}\frac{\beta \overline{i}^{+}}{\overline{s}%
^{+}+\overline{i}^{+}+\overline{r}^{+}+\varepsilon }\,s(s-s_{M})^{+}dxd\tau
\leq 0
\end{eqnarray*}%
and can infer that $s\leq s_{M}$ a.e.~in $Q$. Thus, we have proved that 
\begin{equation}
0\leq s\leq s_{M}\text{ a.e.~in }Q.  \label{s-bound}
\end{equation}%
Next, we deal with (\ref{eq-i2}) rewritten as 
\begin{equation}
\partial _{t}i-\nabla \cdot (\nu _{i}\nabla i)+\phi i-\frac{\beta \overline{s%
}^{+}}{\overline{s}^{+}+\overline{i}^{+}+\overline{r}^{+}+\varepsilon }%
\,i=-\nabla \cdot (\chi (\overline{i}^{+})\nabla s),  \label{eq-i3}
\end{equation}%
together with $\nabla i\cdot \mathbf{n}=0$ on $\Sigma $ and $i(0)=i_{0}.$ We
note that 
\begin{equation}
0\leq \frac{\beta \overline{s}^{+}}{\overline{s}^{+}+\overline{i}^{+}+%
\overline{r}^{+}+\varepsilon }\leq \beta _{M}\text{ \ a.e.~in }Q
\label{beta-bound1}
\end{equation}%
and that the right-hand side of (\ref{eq-i3}) belongs to $%
L^{2}(0,T;V^{\prime })$ thanks to (\ref{chi2}). Relying again on the
existence results for parabolic problems, we deduce that there exists a
unique solution 
\begin{equation}
i\in H^{1}(0,T;V^{\prime })\cap C([0,T];H)\cap L^{2}(0,T;V)  \label{i-reg}
\end{equation}%
to the problem (\ref{eq-i3}) along with $\nabla i\cdot \mathbf{n}=0$ on $%
\Sigma $ and $i(0)=i_{0}$.

As $i$ is now fixed, applying the same argument as before yields that (\ref%
{eq-r2}), together with $\nabla r\cdot \mathbf{n}=0$ on $\Sigma $ and $%
r(0)=r_{0}$, has a unique solution 
\begin{equation}
r\in H^{1}(0,T;V^{\prime })\cap C([0,T];H)\cap L^{2}(0,T;V).  \label{r-reg}
\end{equation}

Next, we deduce some uniform estimates. In the sequel, we shall denote by $C$
several positive constants (that may differ from a line to the other)
depending on the problem parameters, but independent of $\varepsilon .$

First, we test (\ref{eq-s2}) by $s(t),$ integrate over $(0,t),$ taking (\ref{niu}) and (\ref{beta}) \pier{into account}, and obtain%
\[
\frac{1}{2}\left\Vert s(t)\right\Vert _{H}^{2}+\nu
_{m}\int_{0}^{t}\left\Vert \nabla s(\tau )\right\Vert _{H}^{2}d\tau \leq 
\frac{1}{2}\left\Vert s_{0}\right\Vert _{H}^{2}+\beta
_{M}\int_{0}^{t}\left\Vert s(\tau )\right\Vert _{H}^{2}d\tau , 
\]%
whence, by the Gronwall lemma, we deduce the estimate 
\begin{equation}
\left\Vert s\right\Vert _{L^{\infty }(0,T;H)\cap L^{2}(0,T;V)}\leq C,
\label{est-s}
\end{equation}%
where $C$ depends only on $\left\Vert s_{0}\right\Vert _{H},$ $\nu _{m},$ $%
\beta _{M}$ and $T.$ Further, by the comparison of the terms in (\ref{eq-s2}%
) we also infer that 
\begin{equation}
\left\Vert \partial _{t}s\right\Vert _{L^{2}(0,T;V^{\prime })}\leq C.
\label{est-s1}
\end{equation}%
Next, by testing (\ref{eq-i3}) by $i(t)$ and integrating over $(0,t)$ we
have that%
\begin{eqnarray}
&&\frac{1}{2}\left\Vert i(t)\right\Vert _{H}^{2}+\nu
_{m}\int_{0}^{t}\left\Vert \nabla i(\tau )\right\Vert _{H}^{2}d\tau +\phi
\int_{0}^{t}\left\Vert i(\tau )\right\Vert _{H}^{2}d\tau  \nonumber \\
&\pier{\leq}& \frac{1}{2}\left\Vert i_{0}\right\Vert _{H}^{2}+\beta
_{M}\int_{0}^{t}\left\Vert i(\tau )\right\Vert _{H}^{2}d\tau +\chi
_{M}\int_{0}^{t}\left\Vert \nabla s(\tau )\right\Vert _{H}\left\Vert \nabla
i(\tau )\right\Vert _{H}d\tau .  \label{utile3}
\end{eqnarray}
For the last term on the right-hand side, we apply the Young inequality 
\[
\chi_{M}\int_{0}^{t}\left\Vert \nabla s(\tau )\right\Vert _{H}\left\Vert
\nabla i(\tau )\right\Vert _{H}d\tau \leq \frac{\nu _{m}}{2}%
\int_{0}^{t}\left\Vert \nabla i(\tau )\right\Vert _{H}^{2}d\tau +\frac{|\chi
_{M}|^2}{2\nu_m}\int_{0}^{t}\left\Vert \nabla s(\tau )\right\Vert
_{H}^{2}d\tau , 
\]%
(\ref{est-s}) and the Gronwall lemma in order to obtain 
\begin{equation}
\left\Vert i\right\Vert _{L^{\infty }(0,T;H)\cap L^{2}(0,T;V)}\leq C.
\label{est-i}
\end{equation}%
Now, by comparing the terms in (\ref{eq-i3}) we easily arrive at 
\begin{equation}
\left\Vert \partial _{t}i\right\Vert _{L^{2}(0,T;V^{\prime })}\leq C.
\label{est-i1}
\end{equation}%
Proceeding in the same way in (\ref{eq-r2}) with $\nabla r\cdot \mathbf{n}=0$
on $\Sigma $ and $r(0)=r_{0},$ and using (\ref{est-i}) we infer that%
\begin{equation}
\left\Vert r\right\Vert _{L^{\infty }(0,T;H)\cap L^{2}(0,T;V)}+\left\Vert
\partial _{t}r\right\Vert _{L^{2}(0,T;V^{\prime })}\leq C.  \label{est-r}
\end{equation}%
Due to the estimates (\ref{est-s}), (\ref{est-i}), (\ref{est-r}) it turns
out that there exists a constant $R$ such that 
\[
\left\Vert s\right\Vert _{L^{2}(Q)}+\left\Vert i\right\Vert
_{L^{2}(Q)}+\left\Vert r\right\Vert _{L^{2}(Q)}\leq R 
\]%
and we employ exactly this constant in the definition (\ref{M}) of $M.$

Therefore, it follows that $\Phi (M)\subset M.$ Recalling also (\ref{est-s1}%
) and (\ref{est-i1}), due to the compact embedding of $H^{1}(0,T;V^{\prime
})\cap L^{2}(0,T;V)$ into $L^{2}(0,T;H)\equiv L^{2}(Q)$ \pier{(see, e.g., \cite[p. 58]{Lions-69})}, 
we conclude that $\Phi (M)$ is a compact set of $X.$

It remains to prove that $\Phi $ is continuous. We consider a sequence $(%
\overline{s}_{n},\overline{i}_{n},\overline{r}_{n})\in M$ such that $(%
\overline{s}_{n},\overline{i}_{n},\overline{r}_{n})\rightarrow (\overline{s},%
\overline{i},\overline{r})$ strongly in $X,$ as $n\rightarrow \infty .$
Letting now $(s_{n},i_{n},r_{n})=\Phi (\overline{s}_{n},\overline{i}_{n},%
\overline{r}_{n})$ and $(s,i,r)=\Phi (\overline{s},\overline{i},\overline{r}%
),$ we want to prove that $(s_{n},i_{n},r_{n})\rightarrow (s,i,r)$ strongly
in $X,$ as $n\rightarrow \infty .$

Relying on the previous estimates for the solution to (\ref{eq-s2})-(\ref%
{eq-r2}) we have that 
\begin{eqnarray*}
&&\left\Vert s_{n}\right\Vert _{H^{1}(0,T;V^{\prime })\cap L^{\infty
}(0,T;H)\cap L^{2}(0,T;V)}+\left\Vert i_{n}\right\Vert _{H^{1}(0,T;V^{\prime
})\cap L^{\infty }(0,T;H)\cap L^{2}(0,T;V)} \\
&&+\left\Vert r_{n}\right\Vert _{H^{1}(0,T;V^{\prime })\cap L^{\infty
}(0,T;H)\cap L^{2}(0,T;V)}\leq C.
\end{eqnarray*}%
Hence, by compactness, there is a subsequence of $n,$ still denoted by $n,$
such that%
\begin{eqnarray}
&&s_{n}\rightarrow s,\text{ }i_{n}\rightarrow i,\text{ }r_{n}\rightarrow r%
\text{ }  \nonumber \\
&&\text{weakly in }H^{1}(0,T;V^{\prime })\cap L^{2}(0,T;V),\text{ weak* in }%
L^{\infty }(0,T;H),  \nonumber \\
&&\text{strongly in }L^{2}(0,T;H)\text{ and a.e.~in }Q,\text{ \ as }%
n\rightarrow \infty .  \label{conv}
\end{eqnarray}%
We have to prove that $(s,i,r)$ is exactly $\Phi (\overline{s},\overline{i},%
\overline{r})$ and that the whole sequence $(s_{n},i_{n},r_{n})$ converges
in $X.$ For this aim, we should be able to pass to the limit as $%
n\rightarrow \infty $ in the variational formulation of the problem. Namely,
we have that 
\begin{equation}
\int_{0}^{T}\left\langle \partial _{t}s_{n}(t),v(t)\right\rangle
dt+\int_{Q}\nu _{s}\nabla s_{n}\cdot \nabla vdxdt+\int_{Q}\frac{\beta 
\overline{i}_{n}^{+}}{\overline{s}_{n}^{+}+\overline{i}_{n}^{+}+\overline{r}%
_{n}^{+}+\varepsilon }\,s_{n}vdxdt=0,  \label{sn}
\end{equation}%
\begin{eqnarray}
&&\int_{0}^{T}\left\langle \partial _{t}i_{n}(t),v(t)\right\rangle
dt+\int_{Q}\nu _{i}\nabla i_{n}\cdot \nabla vdxdt+\int_{Q}\phi i_{n}vdxdt 
\nonumber \\
&&-\int_{Q}\frac{\beta \overline{s}_{n}^{+}}{\overline{s}_{n}^{+}+\overline{i%
}_{n}^{+}+\overline{r}_{n}^{+}+\varepsilon }\,i_{n}vdxdt=\int_{Q}\chi (%
\overline{i}_{n}^{+})\nabla s_{n}\cdot \nabla vdxdt,  \label{in}
\end{eqnarray}%
\begin{equation}
\int_{0}^{T}\left\langle \partial _{t}r_{n}(t),v(t)\right\rangle
dt+\int_{Q}\nu _{r}\nabla r_{n}\cdot \nabla vdxdt-\int_{Q}\phi i_{n}vdxdt=0,
\label{rn}
\end{equation}%
for all $v\in L^{2}(0,T;V).$

First, we discuss the convergence of some terms. We have 
\[
\frac{\beta \overline{i}_{n}^{+}}{\overline{s}_{n}^{+}+\overline{i}_{n}^{+}+%
\overline{r}_{n}^{+}+\varepsilon }\,v\rightarrow \frac{\beta \overline{i}^{+}%
}{\overline{s}^{+}+\overline{i}^{+}+\overline{r}^{+}+\varepsilon }\,v\text{
\ a.e.~in }Q 
\]%
and 
\[
\left\vert \frac{\beta \overline{i}_{n}^{+}}{\overline{s}_{n}^{+}+\overline{i%
}_{n}^{+}+\overline{r}_{n}^{+}+\varepsilon }\,v\right\vert \leq \beta
_{M}\left\vert v\right\vert \pier{,}
\]%
\pier{whence}, by the Lebesgue dominated convergence theorem, we deduce that 
\[
\frac{\beta \overline{i}_{n}^{+}}{\overline{s}_{n}^{+}+\overline{i}_{n}^{+}+%
\overline{r}_{n}^{+}+\varepsilon }\,v\rightarrow \frac{\beta \overline{i}^{+}%
}{\overline{s}^{+}+\overline{i}^{+}+\overline{r}^{+}+\varepsilon }\,v\text{
\ strongly in }L^{2}(Q). 
\]%
Recalling (\ref{conv}), as $s_{n}\rightarrow s$ strongly in $L^{2}(Q),$ we
obtain 
\[
\int_{Q}\frac{\beta \overline{i}_{n}^{+}}{\overline{s}_{n}^{+}+\overline{i}%
_{n}^{+}+\overline{r}_{n}^{+}+\varepsilon }\,s_{n}vdxdt\rightarrow \int_{Q}%
\frac{\beta \overline{i}^{+}}{\overline{s}^{+}+\overline{i}^{+}+\overline{r}%
^{+}+\varepsilon }\,svdxdt\text{ as }n\rightarrow \infty . 
\]%
Similarly\pier{,} we deduce that 
\[
\int_{Q}\frac{\beta \overline{s}_{n}^{+}}{\overline{s}_{n}^{+}+\overline{i}%
_{n}^{+}+\overline{r}_{n}^{+}+\varepsilon }\,i_{n}vdxdt\rightarrow \int_{Q}%
\frac{\beta \overline{s}^{+}}{\overline{s}^{+}+\overline{i}^{+}+\overline{r}%
^{+}+\varepsilon }\,ivdxdt\text{ as }n\rightarrow \infty . 
\]%
Still in the second equation we have the convergence of the term on the
right-hand side%
\[
\int_{Q}\chi (\overline{i}_{n}^{+})\nabla s_{n}\cdot \nabla vdxdt\rightarrow
\int_{Q}\chi (\overline{i}^{+})\nabla s\cdot \nabla vdxdt\text{ as }%
n\rightarrow \infty 
\]%
because of the same argument{: indeed}, we have 
\[
\chi (\overline{i}_{n}^{+})\nabla v\rightarrow \chi (\overline{i}^{+})\nabla
v\text{ \ strongly in }(L^{2}(Q))^{d}, 
\]%
due to (\ref{chi1})-(\ref{chi2}) and the Lebesgue dominated convergence
theorem, and 
\[
\nabla s_{n}\rightarrow \nabla s\text{ \ weakly in }(L^{2}(Q))^{d}, 
\]%
due to (\ref{conv}). Then, by passing to the limit in (\ref{sn})-(\ref{rn})
and using again (\ref{conv}) we finally deduce that (\betti{for all $v\in L^2(0,T;V)$})
\begin{equation}
\int_{0}^{T}\left\langle \partial _{t}s(t),v(t)\right\rangle dt+\int_{Q}\nu
_{s}\nabla s\cdot \nabla vdxdt+\int_{Q}\frac{\beta \overline{i}^{+}}{%
\overline{s}^{+}+\overline{i}^{+}+\overline{r}^{+}+\varepsilon }\,svdxdt=0,
\label{s}
\end{equation}%
\begin{eqnarray}
&&\int_{0}^{T}\left\langle \partial _{t}i(t),v(t)\right\rangle
dt+\int_{Q}\nu _{i}\nabla i\cdot \nabla vdxdt+\int_{Q}\phi ivdxdt  \nonumber
\\
&&-\int_{Q}\frac{\beta \overline{s}^{+}}{\overline{s}^{+}+\overline{i}^{+}+%
\overline{r}^{+}+\varepsilon }\,\betti{i}vdxdt=\int_{Q}\chi (\overline{i}%
^{+})\nabla s\cdot \nabla vdxdt,  \label{i}
\end{eqnarray}%
\begin{equation}
\int_{0}^{T}\left\langle \partial _{t}r(t),v(t)\right\rangle dt+\int_{Q}\nu
_{r}\nabla r\cdot \nabla vdxdt-\int_{Q}\phi ivdxdt=0,  \label{r}
\end{equation}%
whence we see that $(s,i,r)$ is the unique solution to the initial boundary
value problem for (\ref{eq-s2})-(\ref{eq-r2}). Moreover, {owing to the lower
semicontinuity of norms,} we note that $(s,i,r)$ satisfies the estimate 
\begin{eqnarray}
&&\left\Vert s\right\Vert _{H^{1}(0,T;V^{\prime })\cap L^{\infty
}(0,T;H)\cap L^{2}(0,T;V)}+\left\Vert i\right\Vert _{H^{1}(0,T;V^{\prime
})\cap L^{\infty }(0,T;H)\cap L^{2}(0,T;V)}  \nonumber \\
&&+\left\Vert r\right\Vert _{H^{1}(0,T;V^{\prime })\cap L^{\infty
}(0,T;H)\cap L^{2}(0,T;V)}\leq C.  \label{sol-bound}
\end{eqnarray}

As the limit is {uniquely determined}, it turns out that the whole sequence $%
(s_{n},i_{n},r_{n})$ converges strongly in $X.$ {In fact, by} contradiction
let us assume that there is a subsequence $n_{k}$ such that $%
(s_{n_{k}},i_{n_{k}},r_{n_{k}})$ does not converge. Hence, by arguing as in (%
\ref{conv}) there exists another subsequence $n_{k,j}$ such that $%
(s_{n_{k,j}},i_{n_{k,j}},r_{n_{k,j}})$ satisfies (\ref{conv}) and
consequently converges to the unique solution $(s,i,r)$ to the initial
boundary value problem for (\ref{eq-s2})-(\ref{eq-r2}). \pier{Then} we arrive at a
contradiction.

Therefore, by the Schauder fixed point theorem, we have proved that there
exists a triplet $(s_{\varepsilon },i_{\varepsilon },r_{\varepsilon })$ that
is a weak solution to system (\ref{eq-s1})-(\ref{ci1}).

{At this point, as $s_{\varepsilon } $ clearly satisfies (\ref{s-bound}),}
we have to show that $i_{\varepsilon }$ and $r_{\varepsilon }$ are
nonnegative in order to complete the proof of (\ref{sol-eps-pos}). To this
end we first introduce a Lipschitz continuous approximation of the logarithm
function. Namely, let $\delta \in (0,1)$ and define 
\[
j_{\delta }(v):=\left\{ 
\begin{array}{l}
\ln \delta , \\ 
\ln v, \\ 
\ln \frac{1}{\delta },%
\end{array}%
\begin{array}{l}
\text{if } v<\delta \\ 
\text{if }\delta \leq v\leq 1/\delta \\ 
\text{if }\pier{v}>1/\delta%
\end{array}%
\right. \text{ for }v\in \mathbb{R}. 
\]%
We note that $j_{\delta }$ is the derivative of the convex and nonnegative
function%
\[
\widehat{j_{\delta }}(v):=\left\{ 
\begin{array}{l}
v\ln \delta -\delta +1, \\ 
v\ln v-v+1, \\ 
v\ln \frac{1}{\delta }-\frac{1}{\delta }+1,%
\end{array}%
\begin{array}{l}
\text{if } v<\delta \\ 
\text{if }\delta \leq v\leq 1/\delta \\ 
\text{if }\pier{v}>1/\delta%
\end{array}%
\right. \text{ for }v\in \mathbb{R}, 
\]%
with $\widehat{j_{\delta }}$ taking the minimum value $0$ at $v=1$.

In the next computations we shall use, for simplicity, the notation $i$ and $%
r$ instead of $i_{\varepsilon }$ and $r_{\varepsilon }.$ We test (\ref{eq-i2}%
) by $j_{\delta }(i)$ and integrate with respect to time. Then, with the
help of (\ref{niu}) we easily have that%
\begin{eqnarray}
&&\int_{\Omega }\widehat{j_{\delta }}(i(t))dx+\int_{Q_{t}\cap \{\delta \leq
i\leq 1/\delta \}}\nu _{m}\frac{\left\vert \nabla i\right\vert ^{2}}{i}%
dxd\tau  \nonumber \\
&\pier{\leq}& \int_{\Omega }\widehat{j_{\delta }}(i_{0})dx+\int_{Q_{t}} \left(\frac{%
\beta \overline{s}^{+}}{\overline{s}^{+}+\overline{i}^{+}+\overline{r}%
^{+}+\varepsilon }-\phi\right) ij_{\delta }(i)dxd\tau  \nonumber \\
&&{}+\int_{Q_{t}\cap \{\delta \leq i\leq 1/\delta \}}\frac{\chi (i^{+})%
}{i}\nabla s\cdot \nabla idxd\tau ,  \label{jcap}
\end{eqnarray}%
where $Q_{t}:=\Omega \times (0,t),$ $t\in (0,T)$. Now, we treat in detail
the terms on the right-hand side. We note that 
\begin{eqnarray*}
&&\int_{\Omega }\widehat{j_{\delta }}(i_{0})dx = \int_{\Omega \cap
\{i_{0}<\delta \}}(i_{0}\ln \delta -\delta +1)dx \\
&&{}+\int_{\Omega \cap \{\delta \leq i_{0}\leq 1/\delta \}}(i_{0}\ln
i_{0}-i_{0}+1)dx +\int_{\Omega \cap \{i_{0}>1/\delta \}}\left( i_{0}\ln 
\frac{1}{\delta }-\frac{1}{\delta }+1\right) dx.
\end{eqnarray*}%
As $i_{0}\geq 0$ a.e.~in $\Omega$ and $\delta <1$, it is clear that 
\[
\int_{\Omega \cap \{i_{0}<\delta \}}(i_{0}\ln \delta -\delta +1)dx \leq
\int_{\Omega} 1dx. 
\]%
For the second integral on the right-hand side above, the reader can easily
check that 
\[
\int_{\Omega \cap \{\delta \leq i_{0}\leq 1/\delta \}}(i_{0}\ln
i_{0}-i_{0}+1)dx\leq \int_{\Omega }(|i_{0}|^{2}+1)dx\leq C, 
\]%
since $i_{0}\in H.$ For the third integral we have that%
\[
\int_{\Omega \cap \{i_{0}>1/\delta \}}\left( i_{0}\ln \frac{1}{\delta }-%
\frac{1}{\delta }+1\right) dx\leq \int_{\Omega \cap \{i_{0}>1/\delta
\}}i_{0}\ln i_{0}dx\leq C. 
\]

About the second term on the right-hand side of (\ref{jcap}), we split the
integral in three parts as for the previous one and discuss each of them
separately. Taking into account the sign of factors and using (\ref{beta}),
we infer that 
\begin{eqnarray*}
&&\int_{Q_{t}\cap \{i<\delta \}}\left(\frac{\beta \overline{s}^{+}}{%
\overline{s}^{+}+\overline{i}^{+}+\overline{r}^{+}+\varepsilon }-\phi\right)
ij_{\delta }(i)dxd\tau \\
&=&\int_{Q_{t}\cap \{i<\delta \}}\left(\frac{\beta \overline{s}^{+}}{%
\overline{s}^{+}+\overline{i}^{+}+\overline{r}^{+}+\varepsilon }-\phi\right)
i \ln \delta dxd\tau \\
&=&\int_{Q_{t}\cap \{i<\delta \}}\left(\frac{\beta \overline{s}^{+}}{%
\overline{s}^{+}+\overline{i}^{+}+\overline{r}^{+}+\varepsilon }-\phi\right)(%
\widehat{j_{\delta }}(i)+\delta -1)dxd\tau \\
&\leq& \beta _{M}\int_{0}^{t}\!\!\int_{\Omega}\widehat{j_{\delta }}(i(\tau
))dxd\tau +\phi \int_{Q_{t}}1 dxd\tau .
\end{eqnarray*}%
Next, we have that%
\begin{eqnarray*}
&&\int_{Q_{t}\cap \{\delta \leq i\leq 1/\delta \}}\left(\frac{\beta 
\overline{s}^{+}}{\overline{s}^{+}+\overline{i}^{+}+\overline{r}%
^{+}+\varepsilon }-\phi\right) ij_{\delta }(i)dxd\tau \\
&=&\int_{Q_{t}\cap \{\delta \leq i\leq 1/\delta \}}\left(\frac{\beta 
\overline{s}^{+}}{\overline{s}^{+}+\overline{i}^{+}+\overline{r}%
^{+}+\varepsilon }-\phi\right)(\widehat{j_{\delta }}(i)+i -1)dxd\tau \\
&\leq& \beta _{M}\int_{0}^{t}\!\!\int_{\Omega}\widehat{j_{\delta }}(i(\tau
))dxd\tau +(\beta _{M}+\phi )\int_{Q_{t}}(\left\vert i\right\vert +1)dxd\tau.
\end{eqnarray*}%
Finally, it turns out that 
\begin{eqnarray*}
&&\int_{Q_{t}\cap \{i>1/\delta \}}\left(\frac{\beta \overline{s}^{+}}{%
\overline{s}^{+}+\overline{i}^{+}+\overline{r}^{+}+\varepsilon }-\phi\right)
ij_{\delta }(i)dxd\tau \\
&=&\int_{Q_{t}\cap \{i>1/\delta \}}\left(\frac{\beta \overline{s}^{+}}{%
\overline{s}^{+}+\overline{i}^{+}+\overline{r}^{+}+\varepsilon }%
-\phi\right)\!\!\left(\widehat{j_{\delta }}(i)+\frac1{\delta}
-1\right)dxd\tau \\
&\leq& \beta _{M}\int_{0}^{t}\!\!\int_{\Omega}\widehat{j_{\delta }}(i(\tau
))dxd\tau + \beta _{M} \int_{Q_{t}}(\left\vert i\right\vert +1)dxd\tau,
\end{eqnarray*}%
due to $1/\delta <i$ \pier{in the set of integration.}

Now we discuss the last term on the right-hand side of (\ref{jcap}). As $%
\chi (i^+) = \chi (i)$ whenever $i\geq 0$, using (\ref{chi2}) and (\ref%
{sol-bound}) we deduce that 
\begin{eqnarray*}
&&\int_{Q_{t}\cap \{\delta \leq i\leq 1/\delta \}}\frac{\chi (i^+)}{i}\nabla
s\cdot \nabla idxd\tau \leq \chi _{1,M}\int_{Q}\left\vert \nabla
s\right\vert \left\vert \nabla i\right\vert dxd\tau \leq C.
\end{eqnarray*}

Then, collecting the previous estimates to control the right hand side of (%
\ref{jcap}), we can apply the Gronwall lemma in order to obtain 
\begin{equation}
\int_{\Omega }\widehat{j_{\delta }}(i(t))dx+\int_{Q_{t}\cap \{\delta \leq
i\leq 1/\delta \}}\nu _{m}\frac{\left\vert \nabla i\right\vert ^{2}}{i}%
dxd\tau \leq C \pier{,} \label{delta-bound}
\end{equation}%
with $C$ independent of $\delta $ and $\varepsilon $ as well. Our aim is now
to pass to the limit as $\delta \to 0$. Since it holds that 
\[
\widehat{j_{\delta }}(i)\rightarrow \widehat{j}(i):=\left\{ 
\begin{array}{l}
+\infty \\ 
1 \\ 
i\ln i-i+1%
\end{array}%
\begin{array}{l}
\text{if }\ i<0 \\ 
\text{if }\ i=0 \\ 
\text{if }\ i>0%
\end{array}%
\right. 
\]%
pointwise, by the Fatou lemma we recover that 
\[
\int_{\Omega }\widehat{j}(i(t))dx\leq C\text{ \ for all }t\in \lbrack 0,T] 
\]%
(recall that $i\in C([0,T];H)$), whence necessarily $i\geq 0$ a.e.~in $Q$.
Moreover, still by virtue of (\ref{delta-bound}), we infer that 
\[
\int_{Q}4\nu _{m}\left\vert \nabla i^{1/2}\right\vert ^{2}dxd\tau \leq C, 
\]%
which offers the additional estimate (cf.~(\ref{sol-bound})) 
\begin{equation}
\left\Vert i^{1/2}\right\Vert _{L^{\infty }(0,T;L^{4}(\Omega ))\cap
L^{2}(0,T;V)}\leq C.  \label{sol-bound-2}
\end{equation}%
At this point, it remains to prove that $r$ is non negative. For this aim it
suffices to test (\ref{eq-r1}) by the negative part $r^{-}(t)$, integrate
and use (\ref{ci-cond}) and the sign of the term $-\phi i$ in the equation.
Therefore, we can drop out the positive parts in (\ref{eq-s1})-(\ref{eq-i1})
and conclude the proof of Theorem \ref{t-uno}.\hfill $\square $

\begin{theorem}
\label{t-due} Under the assumptions (\ref{niu})-(\ref{ci-cond}) there exists
a triplet 
\begin{equation}
(s,i,r)\in (H^{1}(0,T;V^{\prime })\cap C([0,T];H)\cap L^{2}(0,T;V))^{3}
\label{sol-fin}
\end{equation}%
that is a weak solution to system (\ref{eq-s})-(\ref{ci}), with the
properties 
\begin{equation}
0\leq s\leq s_{M},\text{ }i\geq 0,\text{ }r\geq 0\text{ a.e.~in }Q
\label{sol-fin-pos}
\end{equation}%
and where the term%
\begin{eqnarray}
&&\frac{\beta si}{s+i+r}\text{ in equations (\ref{eq-s}) and (\ref{eq-i})} 
\nonumber \\
&&\text{has to be understood as }0\text{ whenever }s+i+r=0.  \label{frac}
\end{eqnarray}%
Moreover, if we set $n=s+i+r$ the conservation property%
\begin{equation}
\int_{\Omega }n(x,t)dx=\int_{\Omega }n_{0}(x)dx,\text{ for all }t\in \lbrack
0,T],  \label{cons}
\end{equation}%
holds, where $n_{0}:=s_{0}+i_{0}+r_{0}.$
\end{theorem}

\medskip

\noindent \textbf{Proof. }We recall that the weak solution $(s_{\varepsilon
},i_{\varepsilon },r_{\varepsilon })$ to equations (\ref{eq-s-eps})-(\ref%
{eq-r-eps}), with the boundary conditions (\ref{bc1}) and the initial
conditions (\ref{ci1}), satisfies\ the uniform estimates (\ref{sol-bound}), (%
\ref{sol-bound-2}) in addition to the properties (\ref{sol-eps-pos}). Then,
there exist a triplet $(s,i,r)$ and a subsequence, still denoted by $%
\varepsilon ,$ such that 
\begin{eqnarray}
&&s_{\varepsilon }\rightarrow s,\text{ }i_{\varepsilon }\rightarrow i,\text{ 
}r_{\varepsilon }\rightarrow r\text{ }  \nonumber \\
&&\text{weakly in }H^{1}(0,T;V^{\prime })\cap L^{2}(0,T;V),\text{ weak* in }%
L^{\infty }(0,T;H),  \nonumber \\
&&\text{strongly in }L^{2}(0,T;H)\text{ and a.e.~in }Q,\text{ \ as }%
\varepsilon \rightarrow 0,  \label{conv-fin}
\end{eqnarray}%
with $(s,i,r)$ satisfying (\ref{sol-fin-pos}). Then, we aim to pass to the
limit in the variational form of (\ref{eq-s-eps})-(\ref{eq-r-eps}), that is, 
\[
\int_{0}^{T}\left\langle \partial _{t}s_{\varepsilon }(t),v(t)\right\rangle
dt+\int_{Q}\nu _{s}\nabla s_{\varepsilon }\cdot \nabla vdxdt+\int_{Q}\frac{%
\beta i_{\varepsilon }}{s_{\varepsilon }+i_{\varepsilon }+r_{\varepsilon
}+\varepsilon }\,s_{\varepsilon }vdxdt=0, 
\]%
\begin{eqnarray*}
&&\int_{0}^{T}\left\langle \partial _{t}i_{\varepsilon
}(t),v(t)\right\rangle dt+\int_{Q}\nu _{i}\nabla i_{\varepsilon }\cdot
\nabla vdxdt+\int_{Q}\phi i_{\varepsilon }vdxdt \\
&&-\int_{Q}\frac{\beta s_{\varepsilon }}{s_{\varepsilon }+i_{\varepsilon
}+r_{\varepsilon }+\varepsilon }\,i_{\varepsilon }vdxdt=\int_{Q}\chi
(i_{\varepsilon })\nabla s_{\varepsilon }\cdot \nabla vdxdt,
\end{eqnarray*}%
\[
\int_{0}^{T}\left\langle \partial _{t}r_{\varepsilon }(t),v(t)\right\rangle
dt+\int_{Q}\nu _{r}\nabla r_{\varepsilon }\cdot \nabla vdxdt-\int_{Q}\phi
i_{\varepsilon }vdxdt=0, 
\]%
for all $v\in L^{2}(0,T;V).$ We note that $\frac{\beta i_{\varepsilon }}{%
s_{\varepsilon }+i_{\varepsilon }+r_{\varepsilon }+\varepsilon }$ converges
a.e.~and weak* in $L^{\infty }(Q)$ to a function $g$ which is bounded,
staying between $0$ and $\beta _{M}$, and is equal to $\frac{\beta i}{s+i+r}$
when $s+i+r>0$. On the other hand, since $s_{\varepsilon }\rightarrow s$
a.e., it turns out that the limit product $gs$ should satisfy (\ref{frac}):
indeed, $s=0$ when $s+i+r=0$. The same conclusion holds also for the similar
term in the second variational equality.

For the integral on the right-hand side of the second equation, we have that 
\begin{equation}
\int_{Q}\chi (i_{\varepsilon })\nabla s_{\varepsilon }\cdot \nabla
vdxdt\rightarrow \int_{Q}\chi (i)\nabla s\cdot \nabla vdxdt \, \text{ as }%
\varepsilon \rightarrow 0 \pier{.}
\label{pierX}
\end{equation}
\pier{Indeed, it holds that}
\[
\chi (i_{\varepsilon })\nabla v\rightarrow \chi (i)\nabla v\text{ \ strongly
in }(L^{2}(Q))^{d}, 
\]%
by virtue of (\ref{chi1})-(\ref{chi2}) and the Lebesgue dominated
convergence theorem\pier{; then,} on account of the weak convergence of $\nabla
s_{\varepsilon }$ to $\nabla s$ in $(L^{2}(Q))^{d},$ due to (\ref{conv-fin}), \pier{(\ref{pierX}) follows.}

As for the other integrals in the variational equalities there is no trouble
for passing to the limit, we can conclude that $(s,i,r)$ is a weak solution
to (\ref{eq-s})-(\ref{ci}).

In order to prove (\ref{cons}), we sum up the equations (\ref{eq-s}), (\ref%
{eq-i}), (\ref{eq-r}), then test the sum by $1$. Using the boundary
conditions and integrating from $0 $ to $t$, it is a standard matter to
arrive at (\ref{cons}). This ends the proof.\hfill $\square $

\medskip

Now, under the additional hypothesis that $s_{0}$ is bounded from below by a
positive constant, i.e., 
\begin{equation}
\frac{1}{s_{0}}\in L^{\infty }(\Omega )  \label{s0-bound}
\end{equation}%
we prove the following result.

\begin{proposition}
\label{prop-tre}Let the assumptions of Theorem \ref{t-due} hold. In
addition, we assume (\ref{s0-bound}). Then there exists a constant $s_{m}>0$
such that 
\begin{equation}
s_{m}\leq s\leq s_{M}\text{ \ a.e.~in }Q.  \label{s-boound}
\end{equation}
\end{proposition}

\noindent\textbf{Proof.} We would like to test equation (\ref{eq-s}) by the
function $v=-ps^{-p-1}$, for $p\geq 1,$ then deduce a uniform estimate for $%
\left\Vert \frac{1}{s}\right\Vert _{p}$ and then pass to the limit as $%
p\rightarrow \infty .$ The chosen test function is not in $V,$ so that it
should be replaced by a smooth approximation, as we rigorously did before in
the calculation of the estimate for the function $\widehat{j}(i)$ in the
proof of Theorem~\ref{t-uno}. However, here we shall skip the rigorous
approach and perform only the formal computation. Hence, we test (\ref{eq-s}%
) by the function $v=-ps^{-p-1}$ and \pier{obtain}
\begin{equation}
\frac{d}{d t}\int_{\Omega }\frac{1}{s^{p}}dx+\int_{\Omega }\nu
_{s}p(p+1)s^{-p-2}\left\vert \nabla s\right\vert ^{2}dx=\int_{\Omega }\frac{%
\beta i}{s+i+r}ps^{-p}dx.  \label{s1}
\end{equation}%
We observe that 
\[
\left\vert \frac{\left\vert \nabla s\right\vert }{s^{1+p/2}}\right\vert
^{2}=\left\vert \frac{2}{p}\nabla s^{-p/2}\right\vert ^{2} 
\]%
and integrate (\ref{s1}) from $0$ to $t.$ Using (\ref{niu}) and (\ref{beta})
we obtain%
\begin{eqnarray*}
&&\int_{\Omega }\left\vert \frac{1}{s(t)}\right\vert ^{p}dx+4\nu _{m}\frac{%
p+1}{p}\int_{0}^{t}\!\!\int_{\Omega}\left\vert \nabla s^{-p/2}\right\vert
^{2}dxd\tau \\
&\leq& \int_{\Omega }\left\vert \frac{1}{s_{0}}\right\vert ^{p}dx+\beta
_{M}p\int_{0}^{t}\!\!\int_{\Omega}\left\vert \frac{1}{s(\tau )}\right\vert
^{p}dxd\tau .
\end{eqnarray*}%
By applying the Gronwall lemma we deduce that%
\begin{eqnarray*}
&&\int_{\Omega }\left\vert \frac{1}{s(t)}\right\vert ^{p}dx+4\nu _{m}\frac{%
p+1}{p}\int_{0}^{t}\!\!\int_{\Omega}\left\vert \nabla s^{-p/2}\right\vert
^{2}dxd\tau \\
&\leq & e^{\beta _{M}pT}\int_{\Omega }\left\vert \frac{1}{s_{0}}\right\vert
^{p}dx,\text{ \ for all }t\in \lbrack 0,T],
\end{eqnarray*}
whence we have 
\[
\left\Vert \frac{1}{s(t)}\right\Vert _{L^{p}(\Omega )}\leq e^{\beta
_{M}T}\left\Vert \frac{1}{s_{0}}\right\Vert _{L^{p}(\Omega )},\text{ \ for
all }t\in \lbrack 0,T]. 
\]%
Then we pass to the limit as $p\rightarrow \infty $ and obtain that 
\[
\left\Vert 1/s\right\Vert _{L^\infty(0,T;L^{\infty }(\Omega ))}\leq e^{\beta
_{M}T}\left\Vert 1/s_{0}\right\Vert _{L^{\infty }(\Omega )}, 
\]
whence 
\[
s\geq \frac{e^{-\beta _{M}T}}{\left\Vert 1/s_{0}\right\Vert _{L^{\infty
}(\Omega )}}=:s_{m} \ \text{ a.e.~in }Q. 
\]%
We point out that another consequence of this argument is 
\begin{equation}
\left\Vert s^{-p/2}\right\Vert _{L^{\infty }(0,T;H)\cap L^{2}(0,T;V)}\leq C,%
\text{ for all }p\in \lbrack 1,+\infty ),  \label{s-p}
\end{equation}%
with $C$ independent of $p.$ This concludes the proof.\hfill $\square $

\medskip

The previous result implies that 
\begin{equation}
0<s_{m}\leq n= s+i+r \ \text{ a.e.~in }Q.  \label{n-bound}
\end{equation}

Next, we are going to prove some additional regularity properties for $s$.

\begin{proposition}
\label{prop-quattro}Under the same assumptions as in Theorem \ref{t-due}, we
assume in addition (\ref{s0-bound}) and%
\begin{equation}
\nu _{s}\text{ is constant, }\beta \in L^{2}(0,T;V),\text{ }s_{0}\in W.
\label{nius-ct}
\end{equation}%
Then, the component $s$ of the solution $(s,i,r)$ found by Theorem \ref%
{t-due} is also a strong solution of (\ref{eq-s}) and satisfies 
\begin{equation}
\left\Vert s\right\Vert _{H^{1}(0,T;V)\cap L^{\infty }(0,T;W)}\leq C.
\label{est-Linf}
\end{equation}
\end{proposition}

\noindent\textbf{Proof.} In preparation of the following calculations we set
the notation%
\begin{equation}
\psi (s,i,r)=\frac{si}{s+i+r}  \label{notpsi}
\end{equation}
and, on account of (\ref{sol-fin-pos}), claim that $0\leq \psi (s,i,r)\leq
s_{M}$ a.e.~in $Q.$

We first test (\ref{eq-s}) by $\partial _{t}s$ and integrate with respect to
time. Then, by the Young inequality we obtain
\begin{eqnarray*}
&&\int_{0}^{t}\!\!\int_{\Omega}\left\vert \partial _{t}s\right\vert ^{2}dxd\tau
+\frac{\nu_s}{2}\int_{\Omega }\left\vert \nabla s(t)\right\vert ^{2}dx 
\nonumber \\
&=&\frac{\nu_s}{2}\int_{\Omega }\left\vert \nabla s_{0}\right\vert
^{2}dx-\int_{0}^{t}\!\!\int_{\Omega}\beta \psi (s,i,r)(\partial _{t}s)dxd\tau \\
&\leq& \frac{\nu_s}{2}\left\Vert s_{0}\right\Vert _{V}^{2}+\frac{1}{2}%
\int_{0}^{t}\!\!\int_{\Omega}\left\vert \partial _{t}s\right\vert ^{2}dxd\tau +%
\frac{1}{2}\beta _{M}^{2}s_{M}^{2}T\, \text{meas}(\Omega),
\end{eqnarray*}%
which obviously implies that $\left\Vert s\right\Vert _{H^{1}(0,T;H)\cap
L^{\infty }(0,T;V)}\leq C.$ By a subsequent comparison of terms in (\ref%
{eq-s}) we infer that $\left\Vert \Delta s\right\Vert _{L^{2}(0,T;H)}\leq C$%
, whence by elliptic regularity we find out that%
\begin{equation}
\left\Vert s\right\Vert _{H^{1}(0,T;H)\cap L^{\infty }(0,T;V)\cap
L^{2}(0,T;W)}\leq C.  \label{est-H2}
\end{equation}%
Now we are going to prove (\ref{est-Linf}). We proceed formally by testing (\ref{eq-s}) by $-\Delta \partial _{t}s$ and integrating from $0$ to $t.$ We
have%
\begin{eqnarray}
&&\int_{0}^{t}\!\!\int_{\Omega}\left\vert \nabla \partial _{t}s\right\vert
^{2}dxd\tau +\frac{\nu _{s}}{2}\int_{\Omega }\left\vert \Delta
s(t)\right\vert ^{2}dx  \nonumber \\
&\leq& \frac{\nu _{s}}{2}\int_{\Omega }\left\vert \Delta s_{0}\right\vert
^{2}dx-\int_{0}^{t}\!\!\int_{\Omega}\psi (s,i,r)\nabla \beta \cdot \nabla
(\partial _{t}s)dxd\tau  \nonumber \\
&&{}-\int_{0}^{t}\!\!\int_{\Omega}\beta \nabla \psi (s,i,r)\cdot \nabla
(\partial _{t}s) dxd\tau .  \label{est-inter}
\end{eqnarray}%
In the last two terms on the right-hand side we apply the Young inequality
to absorb the resulting contribution $\frac 1 2 \int_{0}^{t}\int_{\Omega
}\vert \nabla \partial _{t}s\vert^2 $ with the respective term on the
left-hand side. Besides, we observe that 
\begin{equation}
\int_{0}^{t}\!\!\int_{\Omega} \vert \psi (s,i,r)\vert^2 \left\vert \nabla \beta
\right\vert^{2}dxd\tau \leq C  \label{utile1}
\end{equation}
since $|\psi (s,i,r)|\pier{^2}\leq s_{M}^2$ and $\beta \in L^{2}(0,T;V).$ Now, we
calculate 
\begin{equation}
\nabla \psi (s,i,r)=\frac{i(i+r)}{(s+i+r)^{2}}\nabla s+\frac{s(s+r)}{%
(s+i+r)^{2}}\nabla i-\frac{si}{(s+i+r)^{2}}\nabla r  \label{utile2}
\end{equation}
and note that \pier{the absolute value of every fraction is} smaller than $1$, so that \pier{$\nabla \psi (s,i,r)$} is
bounded in $(L^{2}(Q))^{d}$ from (\ref{sol-bound}). In conclusion, by (\ref%
{est-inter}), (\ref{nius-ct}) and (\ref{est-H2}) it is straightforward to
deduce~(\ref{est-Linf}).\hfill $\square $

\medskip

Thanks to the previous results we are now able to prove the uniqueness of
the solution.

\begin{theorem}
\label{t-cinque} Let (\ref{niu})-(\ref{ci-cond}), (\ref{s0-bound}), (\ref%
{nius-ct}) hold and assume in addition that there exists a constant $L_M$
such that 
\begin{equation}
|\chi (x,t,v_1) - \chi(x,t,v_2)| \leq L_M |v_1 - v_2| \ \text{ for all }\,
v_1, v_2 \geq 0, \ \hbox{a.e. } (x,t) \in Q.  \label{chi4}
\end{equation}%
Then, the problem (\ref{eq-s})-(\ref{ci}) has a unique weak solution $%
(s,i,r) $.
\end{theorem}

\medskip

\noindent \textbf{Proof.} Let $(s_{j},i_{j},r_{j}),$ $j=1,2,$ be two weak
solutions to (\ref{eq-s})-(\ref{ci}) and, within this proof, set $%
s:=s_{1}-s_{2},$ $i:=i_{1}-i_{2},$ $r:=r_{1}-r_{2}.$ We \pier{consider} the difference
of the equations (\ref{eq-s}), add the term $\nu _{s}s$ to both sides, then
test by $\partial _{t}s.$ At the same time, we take the difference of the
equations (\ref{eq-i}), testing by $i,$ and the difference of the equations (%
\ref{eq-r}), testing by $r.$ Then we add all the resultants and obtain 
\begin{eqnarray}
&&\int_{Q_{t}}\left\vert \partial _{t}s\right\vert ^{2}dxd\tau +\frac{\nu
_{s}}{2}\left\Vert s(t)\right\Vert _{V}^{2}+\frac{1}{2}\left\Vert
i(t)\right\Vert _{H}^{2}+\nu _{m}\int_{Q_{t}}\left\vert \nabla i\right\vert
^{2}dxd\tau  \nonumber \\
&& {}+\int_{Q_{t}}\phi i^{2}dxd\tau +\frac{1}{2}\left\Vert
r(t)\right\Vert _{H}^{2}+\nu _{m}\int_{Q_{t}}\left\vert \nabla r\right\vert
^{2}dxd\tau  \nonumber \\
&\leq &\int_{Q_{t}}\nu _{s}s\partial _{t}sdxd\tau +\int_{Q_{t}}\beta (\psi
(s_{1},i_{1},r_{1})-\psi (s_{2},i_{2},r_{2}))(i-\partial _{t}s)dxd\tau 
\nonumber \\
&& {}+\int_{Q_{t}}(\chi (i_{1})\nabla s_{1}-\chi (i_{2})\nabla
s_{2})\cdot \nabla idxd\tau +\int_{Q_{t}}\phi irdxd\tau ,  \label{pier1}
\end{eqnarray}%
where \pier{the notation $\psi $ from the previous proof has been used}. Now we treat some
terms. By a direct verification we see that there is a positive constant $%
C_{\psi }$ such that 
\[
\left\vert \psi (s_{1},i_{1},r_{1})-\psi (s_{2},i_{2},r_{2})\right\vert \leq
C_{\psi }(\left\vert s\right\vert +\left\vert i\right\vert +\left\vert
r\right\vert ) 
\]%
and by repeated use of the Young inequality we deduce that 
\begin{eqnarray*}
&&\left\vert \int_{Q_{t}}\nu _{s}s(\partial _{t}s)dxd\tau +\int_{Q_{t}}\phi
irdxd\tau \right. \\
&&\quad \left. +\int_{Q_{t}}\beta (\psi (s_{1},i_{1},r_{1})-\psi
(s_{2},i_{2},r_{2}))(i-\partial _{t}s)dxd\tau \right\vert \\
&\leq &\frac{1}{2}\int_{Q_{t}}\left\vert \partial _{t}s\right\vert
^{2}dxd\tau +C\int_{Q_{t}}(s^{2}+i^{2}+r^{2})dxd\tau .
\end{eqnarray*}%
Then, by (\ref{chi4}) we infer that 
\begin{eqnarray*}
&&\int_{Q_{t}}(\chi (i_{1})\nabla s_{1}-\chi (i_{2})\nabla s_{2})\cdot
\nabla idxd\tau \\
&=&\int_{Q_{t}}(\chi (i_{1})-\chi (i_{2}))\nabla s_{1}\cdot \nabla idxd\tau
+\int_{Q_{t}}\chi (i_{2})\nabla s\cdot \nabla idxd\tau \\
&\leq &L_{M}\int_{0}^{t}\left\Vert i(\tau )\right\Vert _{L^{4}(\Omega
)}\left\Vert \nabla s_{1}(\tau )\right\Vert _{L^{4}(\Omega )}\left\Vert
\nabla i(\tau )\right\Vert _{H}d\tau \\
&& {}+\chi _{M}\int_{0}^{t}\left\Vert \nabla s(\tau )\right\Vert
_{H}\left\Vert \nabla i(\tau )\right\Vert _{H}d\tau .
\end{eqnarray*}%
In the first term we use the bound in \pier{(\ref{est-Linf})} (note that $W\subset
W^{1,4}(\Omega )$ continuously) and the compactness of the embedding $%
V\subset L^{4}(\Omega )$, which, along with the Lions lemma (see, e.g., \pier{\cite[p.~59]{Lions-69}}, give%
\begin{eqnarray*}
&&L_{M}\int_{0}^{t}\left\Vert i(\tau )\right\Vert _{L^{4}(\Omega
)}\left\Vert \nabla s_{1}(\tau )\right\Vert _{L^{4}(\Omega )}\left\Vert
\nabla i(\tau )\right\Vert _{H}d\tau \\
&\leq &\frac{\nu _{m}}{2}\int_{Q_{t}}\left\vert \nabla i\right\vert
^{2}dxd\tau +C\int_{0}^{t}\left\Vert s_{1}\right\Vert _{L^{\infty
}(0,T;W)}^{2}\left\Vert i(\tau )\right\Vert _{L^{4}(\Omega )}^{2}d\tau \\
&\leq &\frac{\nu _{m}}{2}\int_{Q_{t}}\left\vert \nabla i\right\vert
^{2}dxd\tau +\delta \int_{0}^{t}\left\Vert i(\tau )\right\Vert _{V}^{2}d\tau
+C_{\delta }\int_{0}^{t}\left\Vert i(\tau )\right\Vert _{H}^{2}d\tau
\end{eqnarray*}%
for all $\delta >0$ with a related constant $C_{\delta }$. Letting $\delta
<\pier{\nu _{m}/2}$, we can collect everything and apply the Gronwall lemma to show
that the left-hand side of (\ref{pier1}) is equal to zero for all $t\in
\lbrack 0,T]$, whence $s=i=r=0$ and the theorem is completely proved. \hfill 
$\square $

\section{Analysis of a reduced system}

\setcounter{equation}{0}

In practice, \pier{let us point out that it makes sense to assume $\nu _{s}$ and $\nu _{r}$ small, and in particular $\nu_{s}, \, \nu_r \, <<\, \nu_i$,} for a few reasons. First, the $s$ and $r$ \pier{populations} do not spread disease. Second, introducing diffusion in these compartments introduces an irreversible process in which the populations move from high-to-low concentration areas, tending to equilibrium over time. This is not realistic as, generally speaking, people's mobility is transient, and they will return to \pier{an} initial starting location. Consequently, over the time-scales relevant in the current work (days/weeks), we do not expect the spatial distribution of the susceptible or recovered populations to vary significantly.
\pier{Accordingly,} it is often appropriate to \pier{directly} set $\nu
_{s}=\nu _{r}=0,$ see \cite{murray2003mathematical}. In view of \pier{these considerations, the asymptotic behavior of our
system (\ref{eq-s})-(\ref{ci}), as the two coefficients $\nu _{s}$ and $\nu_{r}$ tend to $0$, turns out to be worth of investigation.}

\pier{From now on, $\nu _{s},\,\nu _{r}$ are assumed to be constant.}
Our intention is showing that as $\nu _{s},\,\nu _{r}\rightarrow 0$ the
solution to (\ref{eq-s})-(\ref{ci}) converges in some topology to a solution
of the following problem 
\begin{equation}
\partial _{t}s+\frac{\beta i}{s+i+r}s=0,\text{ in }Q,  \label{eq-s-l}
\end{equation}%
\begin{equation}
\partial _{t}i-\nabla \cdot \left( \nu _{i}\nabla i-\chi (i)\nabla s\right)
+\phi i-\frac{\beta s}{s+i+r}i=0,\text{ in }Q,  \label{eq-i-l}
\end{equation}%
\begin{equation}
\partial _{t}r-\phi i=0,\text{ in }Q,  \label{eq-r-l}
\end{equation}%
\begin{equation}
(\nu _{i}\nabla i-\chi (i)\nabla s)\cdot \mathbf{n}=0,\text{ on }\Sigma ,
\label{bc0-l}
\end{equation}%
\begin{equation}
s(0)=s_{0},\,i(0)=i_{0},\,r(0)=r_{0},\text{ in }\Omega .  \label{ci-l}
\end{equation}%
Namely, we are going to prove the following convergence result.

\begin{theorem}
\label{t-sei} Let (\ref{niu})-(\ref{ci-cond}), (\ref{s0-bound}), and 
\begin{equation}
\beta \in L^{2}(0,T;V),\text{ \ }s_{0}\in V,\text{ \ } r_{0} \in V
\label{pier2}
\end{equation}
hold, where $\nu_s$ and $\nu_r$ in (\ref{niu}) are now replaced by two
arbitrary sequences $\nu_{s,n} $ and $\nu_{r,n} $ of positive numbers
monotonically decreasing and converging to $0$ as $n\to \infty.$ Denote by $%
(s_n, i_n, r_n)$ a weak solution \betti{in $L^2(0,T;V')$} to 
\begin{equation}
\partial_{t}s_n-\nabla \cdot (\nu _{s,n}\nabla s_n)+\frac{\beta i_n}{%
s_n+i_n+r_n}\,s_n=0,\text{ in }Q,  \label{eq-s-2}
\end{equation}%
\begin{equation}
\partial _{t}i_n-\nabla \cdot \left( \nu _{i}\nabla i_n-\chi (i_n)\nabla
s_n\right) +\phi i_n-\frac{\beta s_n}{s_n+i_n+r_n}\,i_n=0,\text{ in }Q,
\label{eq-i-2}
\end{equation}%
\begin{equation}
\partial _{t}r_n-\nabla \cdot (\nu _{r,n}\nabla r_n)-\phi i_n=0,\text{ in }Q,
\label{eq-r-2}
\end{equation}%
\begin{equation}
\nabla s_n\cdot \mathbf{n}=(\nu _{i}\nabla i_n-\chi (i_n)\nabla s_n)\cdot 
\mathbf{n}=\nabla r_n\cdot \mathbf{n}=0,\text{ on }\Sigma,  \label{bc0-2}
\end{equation}%
\begin{equation}
s_n(0)=s_{0},\,i_n(0)=i_{0},\,r_n(0)=r_{0},\text{ in }\Omega ,  \label{ci-2}
\end{equation}%
%
whose existence and the regularity properties 
\begin{equation}
s_m \leq s_{n}\leq s_{M},\text{ }i_{n }\geq 0,\text{ }r_{n}\geq 0\text{
a.e.~in }Q,  \label{pier3}
\end{equation}%
are guaranteed by Theorem~\ref{t-due} and Proposition~\ref{prop-tre}. Then,
the estimate 
\begin{eqnarray}
&&\left\Vert s_n\right\Vert _{H^{1}(0,T;H)\cap L^{\infty}(0,T;V)}
+\left\Vert i_n\right\Vert _{H^{1}(0,T;V^{\prime}) \cap L^{\infty
}(0,T;H)\cap L^{2}(0,T;V)}  \nonumber \\
&&+\left\Vert r_n\right\Vert _{H^{1}(0,T;H)\cap L^{\infty}(0,T;V)}\leq C
\label{pier4}
\end{eqnarray}
holds for some constant $C$ independent of $\nu_{s,n}, \, \nu_{r,n}$.
Moreover, there exist a triplet $(s,i,r)$ and a subsequence, still denoted
by $n$, such that 
\begin{eqnarray}
&&s_{n}\rightarrow s, \text{ }r_{n }\rightarrow r\text{ }  \nonumber \\
&&\text{weakly in }H^{1}(0,T;H),\text{ weak* in }L^{\infty }(0,T;V), 
\nonumber \\
&&\text{strongly in }C([0,T];H)\text{ and a.e.~in }Q,  \label{pier5} \\[2mm]
&&\nu_{s,n} s_{n}\rightarrow 0, \text{ } \nu_{r,n} r_{n }\rightarrow 0\text{ 
}  \nonumber \\
&&\text{strongly in }L^2 (0,T;W )\text{ and a.e.~in }Q,  \label{pier6} \\%
[2mm]
&&i_{n }\rightarrow i  \nonumber \\
&&\text{weakly in }H^{1}(0,T;V^{\prime })\cap L^{2}(0,T;V),\text{ weak* in }%
L^{\infty }(0,T;H),  \nonumber \\
&&\text{strongly in }L^{2}(0,T;H)\text{ and a.e.~in }Q,  \label{pier7}
\end{eqnarray}%
as $n\to \infty$, with the limit triplet $(s,i,r)$ being a weak solution to (%
\ref{eq-s-l})-(\ref{ci-l}), \pier{with (\ref{eq-i-l}) to be understood} \betti{in $L^2(0,T;V')$}.
\end{theorem}

\medskip

\noindent \textbf{Proof.} From Theorem~\ref{t-due} it is clear that 
\[
(s_{n},i_{n},r_{n})\in (H^{1}(0,T;V^{\prime })\cap C([0,T];H)\cap
L^{2}(0,T;V))^{3}
\]%
for all $n\in \mathbb{N}$. In addition, as the initial data $s_{0}$ and $%
r_{0}$ are in $V$ (see (\ref{pier2})) and the terms in the equations (\ref%
{eq-s-2}) and (\ref{eq-r-2}), 
\[
\frac{\beta i_{n}}{s_{n}+i_{n}+r_{n}}\,s_{n}\hbox{ and }-\phi i_{n}\ \text{
are (at least) in }\,L^{2}(0,T;H),
\]%
from the parabolic regularity theory (cf., e.g., \cite{Lions}) it turns out
that both 
\[
s_{n},\,r_{n}\,\hbox{ lie in }\,H^{1}(0,T;H)\cap C([0,T];V)\cap L^{2}(0,T;W)
\]%
and strongly solve (\ref{eq-s-2}), (\ref{eq-r-2}). Then, we are allowed to
test (\ref{eq-s-2}) by $s_{n}-\Delta s_{n}$ and (\ref{eq-r-2}) by $r-\Delta
r_{n}$, integrate by parts using the boundary conditions (\ref{bc0-2}) and
integrate from $0$ to $t$. Next, we add the resultants obtaining 
\begin{eqnarray}
&&\frac{1}{2}\Vert s_{n}(t)\Vert _{V}^{2}+\nu _{s,n}\int_{Q_{t}}\left(
|\nabla s_{n}|^{2}+|\Delta s_{n}|^{2}\right) +\int_{Q_{t}}\frac{\beta i_{n}}{%
s_{n}+i_{n}+r_{n}}\,|s_{n}|^{2}  \nonumber \\
&& {}+\frac{1}{2}\Vert r_{n}(t)\Vert _{V}^{2}+\nu
_{r,n}\int_{Q_{t}}\left( |\nabla r_{n}|^{2}+|\Delta r_{n}|^{2}\right)  
\nonumber \\
&\leq &\frac{1}{2}\Vert s_{0}\Vert _{V}^{2}+\frac{1}{2}\Vert r_{0}\Vert
_{V}^{2}-\int_{Q_{t}}\psi (s_{n},i_{n},r_{n})\nabla \beta \cdot \nabla
s_{n}dxd\tau   \nonumber \\
&& {}-\int_{Q_{t}}\beta \nabla \psi (s_{n},i_{n},r_{n})\cdot \nabla
s_{n}dxd\tau +\int_{Q_{t}}\phi \left( i_{n}r_{n}+\nabla i_{n}\cdot \nabla
r_{n}\right) dxd\tau ,\qquad   \label{pier8}
\end{eqnarray}%
for all $t\in \lbrack 0,T],$ where we have used the notation $\psi $ from (%
\ref{notpsi}). Note that all terms on the left-hand side are nonnegative
and, in view of (\ref{pier2}) and (\ref{utile1}), (\ref{utile2}) that can be
repeated here, we easily infer that 
\begin{eqnarray}
&&-\int_{Q_{t}}\psi (s_{n},i_{n},r_{n})\nabla \beta \cdot \nabla
s_{n}dxd\tau -\int_{Q_{t}}\beta \nabla \psi (s_{n},i_{n},r_{n})\cdot \nabla
s_{n}dxd\tau   \nonumber \\
&\leq &\Vert \nabla \beta \Vert _{L^{2}(0,T;H)}^{2}+C\int_{0}^{t}\Vert
\nabla s_{n}(\tau )\Vert _{H}^{2}d\tau   \nonumber \\
&& {}+\beta _{M}\int_{0}^{t}\left( \Vert \nabla i_{n}(\tau )\Vert
_{H}+\Vert \nabla r_{n}(\tau )\Vert _{H}\right) \Vert \nabla s_{n}(\tau
)\Vert _{H}d\tau .  \nonumber
\end{eqnarray}%
On the other hand, we have that 
\[
\int_{Q_{t}}\phi \left( i_{n}r_{n}+\nabla i_{n}\cdot \nabla r_{n}\right)
dxd\tau \leq 2\phi \int_{0}^{t}\Vert \ i_{n}(\tau )\Vert _{V}\Vert
r_{n}(\tau )\Vert _{V}d\tau .
\]%
At this point, we control the right-hand side of (\ref{pier8}) by the last
inequalities and consider the estimate (\ref{utile3}), which can be repeated
here for $i_{n}$ without any modification. We add the two resulting
inequalities and obtain 
\begin{eqnarray}
&&\frac{1}{2}\Vert s_{n}(t)\Vert _{V}^{2}+\nu _{s,n}\int_{Q_{t}}\left(
|\nabla s_{n}|^{2}+|\Delta s_{n}|^{2}\right) +\frac{1}{2}\Vert r_{n}(t)\Vert
_{V}^{2}  \nonumber \\
&& {}+\nu _{r,n}\int_{Q_{t}}\left( |\nabla r_{n}|^{2}+|\Delta
r_{n}|^{2}\right) +\frac{1}{2}\left\Vert i_{n}(t)\right\Vert _{H}^{2}+\nu
_{m}\int_{0}^{t}\left\Vert \nabla i_{n}(\tau )\right\Vert _{H}^{2}d\tau  
\nonumber \\
&\leq &C+C\int_{0}^{t}\Vert \nabla s_{n}(\tau )\Vert _{H}^{2}d\tau +\beta
_{M}\int_{0}^{t}\left( \Vert \nabla i_{n}(\tau )\Vert _{H}+\Vert \nabla
r_{n}(\tau )\Vert _{H}\right) \Vert \nabla s_{n}(\tau )\Vert _{H}d\tau
\qquad   \nonumber \\
&& {}+2\phi \int_{0}^{t}\Vert \ i_{n}(\tau )\Vert _{V}\Vert r_{n}(\tau
)\Vert _{V}d\tau +\beta _{M}\int_{0}^{t}\left\Vert i_{n}(\tau )\right\Vert
_{H}^{2}d\tau   \nonumber \\
&& {}+\chi _{M}\int_{0}^{t}\left\Vert \nabla s_{n}(\tau )\right\Vert
_{H}\left\Vert \nabla i_{n}(\tau )\right\Vert _{H}d\tau ,  \label{pier9}
\end{eqnarray}%
where we warn that now the positive constant $\nu _{m}$ (see (\ref{niu})) is
a bound from below only for the diffusion coefficient $\nu _{i}$. Next, we
use the Young inequality for the terms on the right-hand side of (\ref{pier9}%
), taking care in particular of 
\begin{eqnarray}
&&\beta _{M}\int_{0}^{t}\Vert \nabla i_{n}(\tau )\Vert _{H}\Vert \nabla
s_{n}(\tau )\Vert _{H}d\tau +2\phi \int_{0}^{t}\Vert \ i_{n}(\tau )\Vert
_{V}\Vert r_{n}(\tau )\Vert _{V}d\tau   \nonumber \\
&& {}+\chi _{M}\int_{0}^{t}\left\Vert \nabla s_{n}(\tau )\right\Vert
_{H}\left\Vert \nabla i_{n}(\tau )\right\Vert _{H}d\tau   \nonumber \\
&\leq &\frac{\nu _{m}}{2}\int_{0}^{t}\Vert \nabla i_{n}(\tau )\Vert
_{H}^{2}d\tau +C\int_{0}^{t}\left( \Vert i_{n}(\tau )\Vert _{H}^{2}+\Vert
s_{n}(\tau )\Vert _{V}^{2}+\Vert r_{n}(t)\Vert _{V}^{2}\right) d\tau . 
\nonumber
\end{eqnarray}%
After that, we apply the Gronwall lemma and, due to the elliptic regularity
theory as well, we find out that 
\begin{eqnarray}
&&\left\Vert s_{n}\right\Vert _{L^{\infty }(0,T;V)}+\sqrt{\nu _{s,n}}%
\left\Vert s_{n}\right\Vert _{L^{2}(0,T;W)}+\left\Vert i_{n}\right\Vert
_{L^{\infty }(0,T;H)\cap L^{2}(0,T;V)}  \nonumber \\
&&+\left\Vert r_{n}\right\Vert _{L^{\infty }(0,T;V)}+\sqrt{\nu _{r,n}}%
\left\Vert r_{n}\right\Vert _{L^{2}(0,T;W)}\leq C,  \label{pier10}
\end{eqnarray}%
for some uniform constant $C$. Now, on account of (\ref{pier10}) and by
comparison of terms in equations (\ref{eq-s-2}), (\ref{eq-i-2}), (\ref%
{eq-r-2}) we deduce that 
\[
\left\Vert \partial _{t}s_{n}\right\Vert _{L^{2}(0,T;H)}+\left\Vert \partial
_{t}i_{n}\right\Vert _{L^{2}(0,T;V^{\prime })}+\left\Vert \partial
_{t}r_{n}\right\Vert _{L^{2}(0,T;H)}\leq C,
\]%
which completes the proof of (\ref{pier4}). Having shown these estimates,
all the properties previously derived in the proofs of Theorem~\ref{t-due}
and Proposition~\ref{prop-tre} continue to hold, in particular for every $%
n\in \mathbb{N}$ the function $i_{n}$ is non negative and $s_{n}$ satisfies (%
\ref{s-bound}) with $s_{m},\,s_{M}$ fixed and independent of $\nu
_{s,n},\,\nu _{r,n}$.

Now, thanks to (\ref{pier4}) and (\ref{pier10}) we can conclude that there
are a subsequence of $n$, still denoted by $n$, and a triplet $(s,i,r)$ such
that\pier{,} by weak and weak* compactness, along with the Ascoli theorem, the
convergences (\ref{pier5})-(\ref{pier7}) hold. Moreover, by passing to the
limit in (\ref{eq-s-2})-(\ref{ci-2}), as 
\[
\nabla \cdot (\nu _{s,n}\nabla s_{n})\rightarrow 0,\ \ \nabla \cdot (\nu
_{r,n}\nabla r_{n})\rightarrow 0\ \text{ strongly in $L^{2}(Q)$ as }%
n\rightarrow \infty ,
\]%
and applying the same arguments as in the previous limit procedures we
easily show that $(s,i,r)$ is a weak solution to~(\ref{eq-s-l})-(\ref{ci-l})
and conclude the proof. \hfill \hfill $\square $

\section{Numerical simulations}

\begin{figure}[ht!]
    \centering
    \includegraphics[width=.7\textwidth]{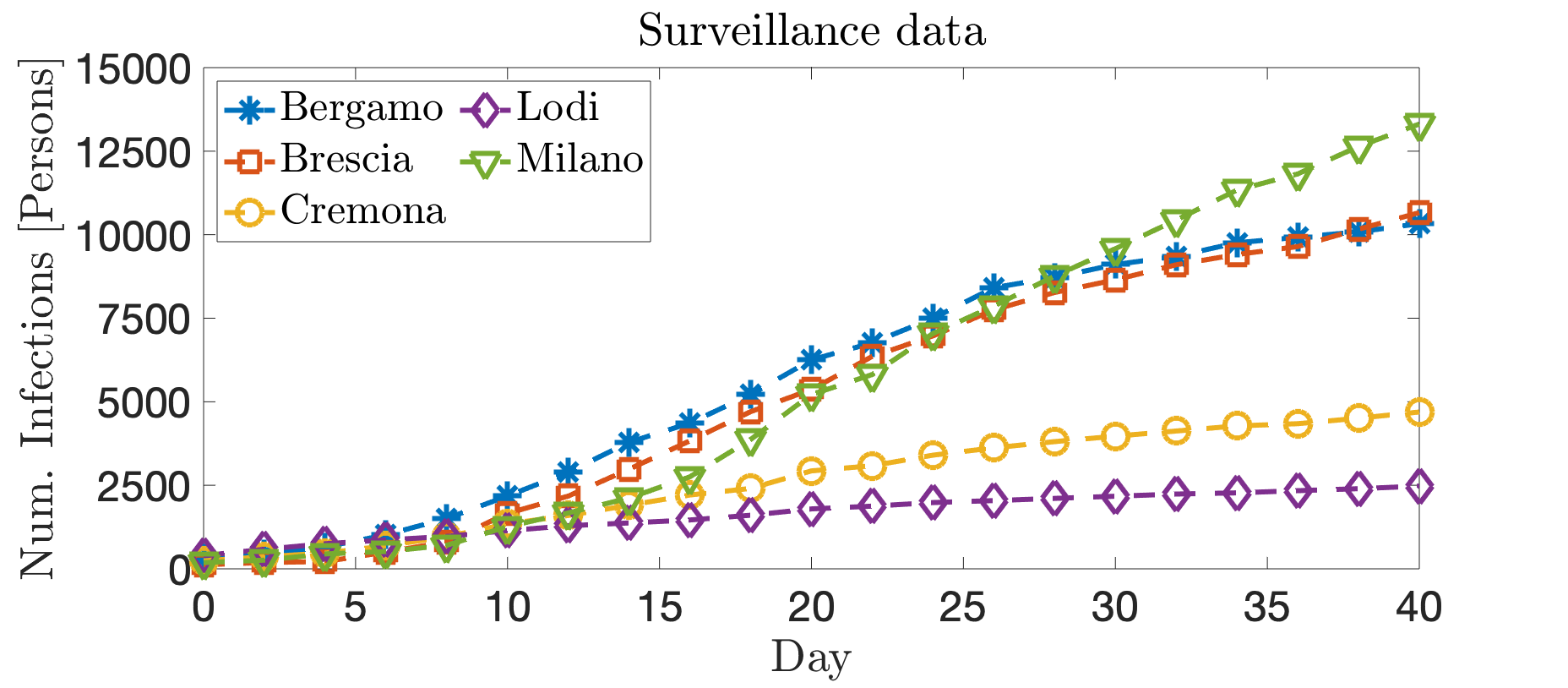}
    \includegraphics[width=.25\textwidth]{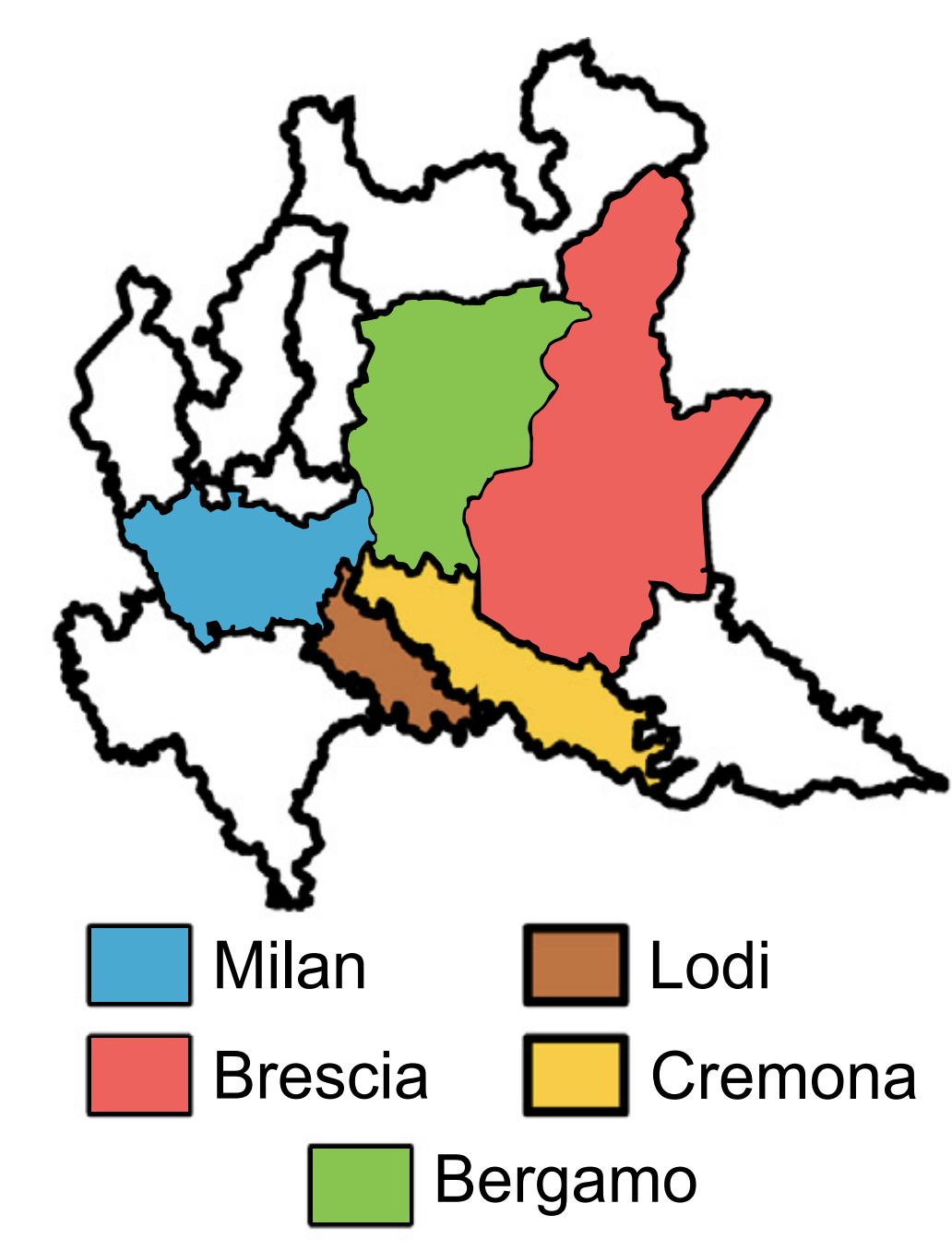}
    \caption{Left: observed cumulative infections across five relevant provinces in Lombardy (data and figure retrived from \cite{Lab24}). Right: geographic locations of each province. }
    \label{fig:map_and_survLomb}
\end{figure}

\setcounter{equation}{0}
In this section, we perform a suite of ten numerical simulations of the reduced model (\ref{eq-s-l})-(\ref{ci-l}) in the Italian region of Lombardy. Across the ten cases, we vary the values of the chemotaxis parameter $\mu_i$, diffusion coefficient $\nu_i$, and capacity term $C_0$ as detailed in Table \ref{tab:simulationParameters}. We are primarily interested in the following basic questions:
\begin{enumerate}
    \item How do changes in $\mu_i$, $\nu_i$ and $C_0$ affect the model behavior, both quantitatively and qualitatively?
    \item Does incorporating a chemotaxis term allow for a more realistic description of airborne infectious disease transmission in human populations? In particular, as mentioned in the introduction, {without resorting to time- and/or space- dependent parameterizations,} is the model able to account for:
    \begin{enumerate}
        \item Transmission over large geographic distances?
        \item The effects of population density and density patterns on transmission?
        \item The relatively rapid appearance of the disease and subsequent sustained transmission in major urban centers, even in the absence of an infected population at initialization?  \end{enumerate}\end{enumerate}
Note that existing PDE models of infectious disease, including those studied in \cite{grave2021assessing, grave2022modeling,bertaglia2021hyperbolic, ramaswamy2021comprehensive}, have difficulty \pier{in} reproducing some or all of these behaviors, {and typically resort to time- and/or space-dependent parameters to do so}.

\par To evaluate the model, we compare the cumulative case counts in the early stages of COVID-19 pandemic with surveillance data. Despite the use of COVID-19 for the example, the introduced model should not be regarded as a model of COVID-19, and the subsequent simulations should not be interpreted as attempts to simulate COVID-19 in a rigorous manner. The choice of this example is based primarily on practical considerations, as the severity, scope, and recency of the COVID-19 pandemic have led to a large amount of publicly available, reasonably high-quality spatiotemporal data. {However, due to well-known difficulties in interpreting such data quantitatively (\cite{bartoszek2020official,backhausCovidData, piazzola-21}), we restrict our comparisons to a qualitative analysis of cumulative incidence in the provinces of Lodi, Cremona, Bergamo, Brescia, and Milano.} A geographic depiction of the relevant locations and the corresponding surveillance data (courtesy of \cite{Lab24}) is provided in Figure \ref{fig:map_and_survLomb}. 
\par The outbreak initially begins in the provinces of Lodi and Cremona (particularly concentrated in the town of Codogno, province of Lodi). However, within the first several weeks, the outbreak becomes far more concentrated in the more populous northern provinces of Bergamo, Brescia and Milan. In our numerical simulations, we hope to observe a similar spatiotemporal dynamics consistent with reality; that is, the initial outbreak in Lodi and Cremona is quickly overshadowed by greater contagion in Bergamo, Brescia, and Milan. Further, ideally we observe these dynamics without changes to the problem parameterization. While updating parameters in order to more precisely account for e.g. lockdowns and other restrictions may be helpful for a detailed model of COVID-19, in the present work we are more interested in simply observing if the chemotaxis model can capture the more general spatio-temporal features observed in reality. Hence, we believe that these core behaviors 2a-2c listed above should be observed, even for a simple, time-constant parameterization.

\par We reconstructed the geometry and defined the initial susceptible and infected compartments following \cite{grave2021assessing}. Regarding the numerical solution, we employed a triangular mesh consisting of 53,506 triangles, giving an average spatial resolution of .44 kilometers. The solution was advanced in time using a BDF2 scheme, using Heun's method at the first time-step to ensure second-order temporal accuracy throughout the simulated time period. We do not discuss the numerical solution of the problem further, however, we note that several relevant considerations regarding the numerical solution for problems of this type can be found in \cite{grave2021assessing, viguerieNumSim}, and apply equally to the current problem.

\begin{table}
\begin{tabularx}{\textwidth}{|X|p{1.35in}|p{.635in}|p{.9in}|p{.92in}|p{.92in}|}
\hline
Sim. & $\mu_i$ & $\nu_i$ & $C_0$ & $\beta$ & $\phi$ \\ \hline\hline
1 &  0.0 $\dfrac{\text{km}^2 \cdot \text{Persons}}{\text{Days}}$ &  1.0 $\dfrac{\text{km}^2}{\text{Days}}$  & 0 Persons  & 0.175 Days$^{-1}$  & 1/18 Days$^{-1}$ \\ \hline

2 &  0.0 $\dfrac{\text{km}^2 \cdot \text{Persons}}{\text{Days}}$ &  2.5 $\dfrac{\text{km}^2}{\text{Days}}$  & 0 Persons  & 0.175 Days$^{-1}$  & 1/18 Days$^{-1}$ \\ \hline

3 &  0.01 $\dfrac{\text{km}^2 \cdot \text{Persons}}{\text{Days}}$ &  1.0 $\dfrac{\text{km}^2}{\text{Days}}$  & 50 Persons  & 0.175 Days$^{-1}$  & 1/18 Days$^{-1}$ \\ \hline

4 &  0.01 $\dfrac{\text{km}^2 \cdot \text{Persons}}{\text{Days}}$ &  1.0 $\dfrac{\text{km}^2}{\text{Days}}$  & 200 Persons  & 0.175 Days$^{-1}$  & 1/18 Days$^{-1}$ \\ \hline

5 &  0.01 $\dfrac{\text{km}^2 \cdot \text{Persons}}{\text{Days}}$ &  2.5 $\dfrac{\text{km}^2}{\text{Days}}$  & 50 Persons  & 0.175 Days$^{-1}$  & 1/18 Days$^{-1}$ \\ \hline

6 &  0.01 $\dfrac{\text{km}^2 \cdot \text{Persons}}{\text{Days}}$ &  2.5 $\dfrac{\text{km}^2}{\text{Days}}$  & 200 Persons  & 0.175 Days$^{-1}$  & 1/18 Days$^{-1}$ \\ \hline

7 &  0.02 $\dfrac{\text{km}^2 \cdot \text{Persons}}{\text{Days}}$ &  1.0 $\dfrac{\text{km}^2}{\text{Days}}$  & 50 Persons  & 0.175 Days$^{-1}$  & 1/18 Days$^{-1}$ \\ \hline

8 &  0.02 $\dfrac{\text{km}^2 \cdot \text{Persons}}{\text{Days}}$ &  1.0 $\dfrac{\text{km}^2}{\text{Days}}$  & 200 Persons  & 0.175 Days$^{-1}$  & 1/18 Days$^{-1}$ \\ \hline

9 &  0.02 $\dfrac{\text{km}^2 \cdot \text{Persons}}{\text{Days}}$ &  2.5 $\dfrac{\text{km}^2}{\text{Days}}$  & 50 Persons  & 0.175 Days$^{-1}$  & 1/18 Days$^{-1}$ \\ \hline

10 &  0.02 $\dfrac{\text{km}^2 \cdot \text{Persons}}{\text{Days}}$ &  2.5 $\dfrac{\text{km}^2}{\text{Days}}$  & 200 Persons  & 0.175 Days$^{-1}$  & 1/18 Days$^{-1}$ \\ \hline

\end{tabularx}%
\caption{Parameter values for each of the ten numerical simulations for Lombardy.}\label{tab:simulationParameters}
\end{table}

\par
\begin{figure}[ht!]
    \centering
    \includegraphics[width=\textwidth]{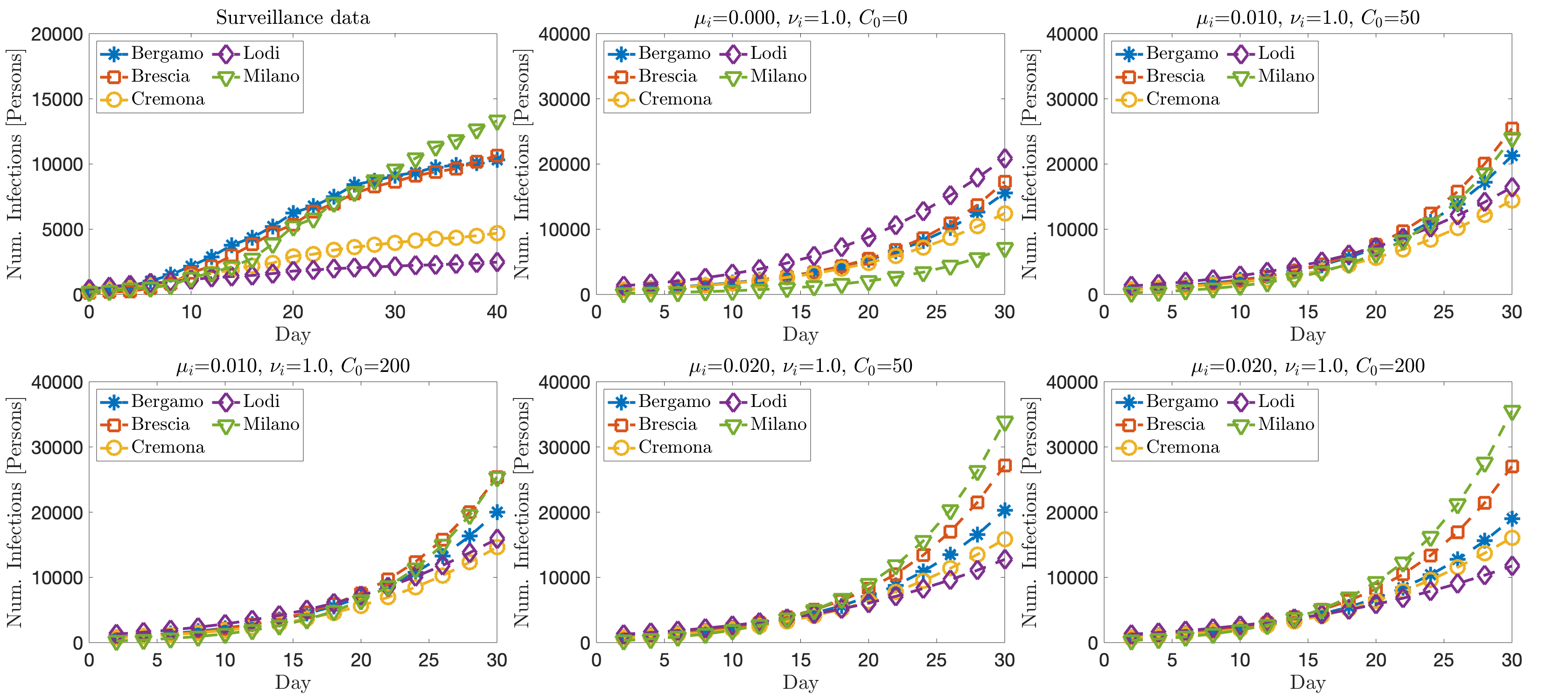}
    \caption{Cumulative simulated incidence in the provinces of Bergamo, Brescia, Lodi, Milano, and Cremona over 30 days, for $\nu_i$=1.0 km$^{2} \cdot $ Days$^{-1}$ and varying values of $\mu_i$ and $C_0$. } 
    \label{fig:lombDiff1pt0}
\end{figure}

\begin{figure}[ht!]
    \centering
    \includegraphics[width=\textwidth]{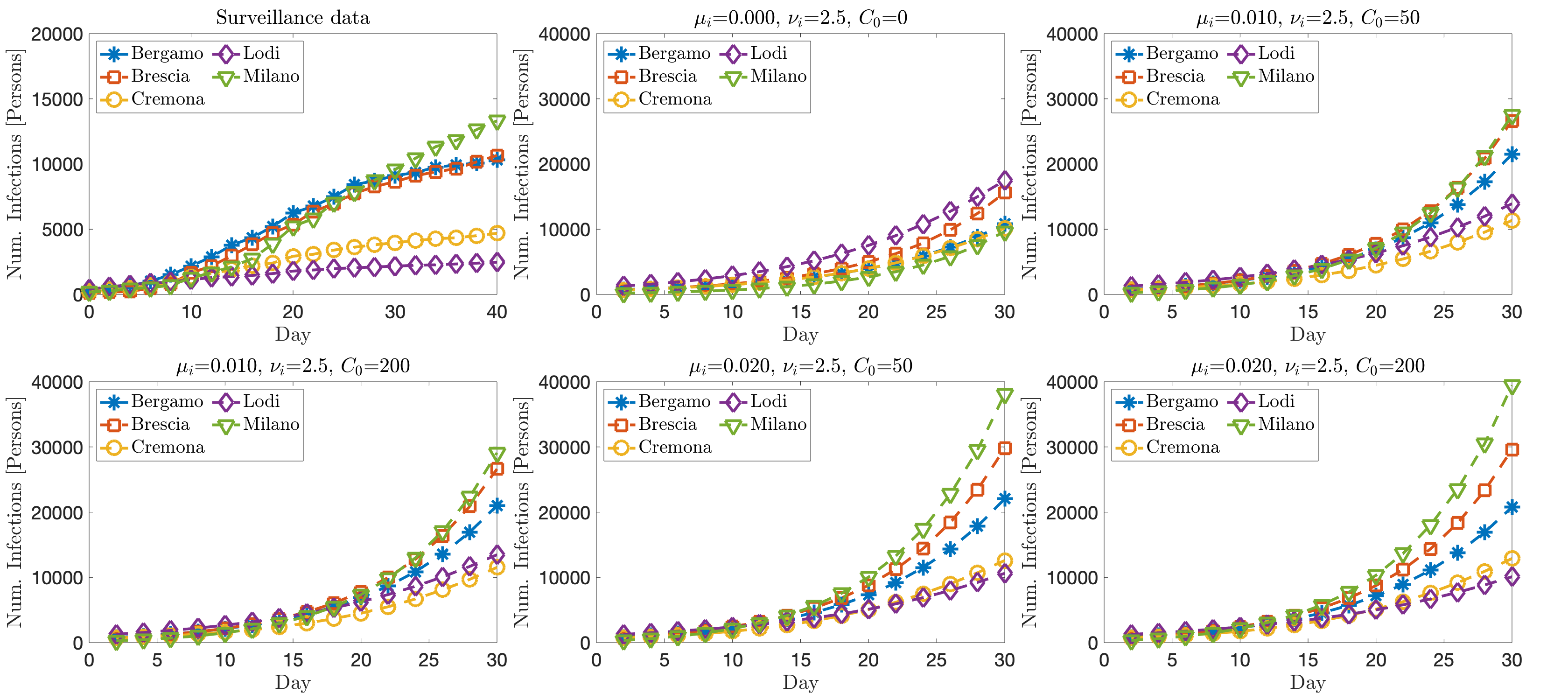}
    \caption{Cumulative simulated incidence in the provinces of Bergamo, Brescia, Lodi, Milano, and Cremona over 30 days, for $\nu_i$=2.5 km$^{2} \cdot $ Days$^{-1}$ and varying values of $\mu_i$ and $C_0$.}
    \label{fig:lombDiff2pt0}
\end{figure}

 The results of the \revis{simulations} are presented in Fig.~\ref{fig:lombDiff1pt0} (for $\nu_i$=1.0 km$^{2}\cdot$Days$^{-1}$) and Fig.~\ref{fig:lombDiff2pt0} (for $\nu_i$=2.5 km$^{2}\cdot$Days$^{-1}$). For both values of $\nu_i$, we see many of the same general trends. In the absence of chemotaxis ($\mu_i=0.0 \, \text{km}^2\cdot\text{Persons}\cdot\text{Days}^{-1}$), the numerical simulations are not qualitatively consistent with the surveillance data. In particular, the outbreak remains most severe in the province of Lodi, and growth is similar in the Milan, Bergamo, and Cremona regions. Once chemotaxis is introduced, the simulations begin to show a better qualitative agreement with the surveillance data. Around day 15, the rate of growth in the Bergamo, Brescia, and Milan regions begins to significantly outpace Cremona and Lodi. {When $\mu_i=.01 \,\text{km}^2\cdot\text{Persons}\cdot\text{Days}^{-1}$, $\nu_i=1.0\, \text{km}^2\cdot\text{Days}^{-1}$, and $C_0=50 \,\text{Persons}$, cumulative cases multiply grow by a factor of 4.1 in Bergamo, 5.7 in Brescia, and 6.3 in Milan from day 15 to day 25; over the same period, cumulative cases increase by a factor of 1.3 in both Cremona and Lodi.} 

Doubling $\mu_i$ from .01 $\text{km}^2\cdot\text{Persons}\cdot\text{Days}^{-1}$ to .02 $\text{km}^2\cdot\text{Persons}\cdot\text{Days}^{-1}$ results in no substantive change in transmission in Brescia or Cremona. However, transmission in Milan increases 34\%, and interestingly, transmission \textit{decreases} by 12\% in Bergamo and by 21\% in Lodi.

We now examine the effect of the diffusion parameter $\nu_i$. In the absence of chemotaxis, increasing diffusion $\nu_i$ from 1.0 $\text{km}^2\cdot\text{Days}^{-1}$ to 2.5 $\text{km}^2\cdot\text{Days}^{-1}$ reduces transmission in most provinces, with cumulative cases decreasing 34\%, 11\%, 10\% and 8\% in Bergamo, Brescia, Cremona, and Lodi, respectively. In Milan, however, increased diffusion, even absent chemotaxis effects, increases transmission by 48\%. This makes sense, as Milan is significantly further away from the initial hotspots; hence, increasing the diffusion leads to the infection front reaching Milan more quickly. However, in the other areas, diffusion primarily serves to reduce the local infection concentration, and hence overall transmission. 

{When chemotaxis is added, however, the effect of diffusion becomes more complex. Increasing diffusion may  increase or decrease transmission, depending on the specific situation. In our simulations, we observed that when $\nu_i$ is increased from {1.0 $\text{km}^2\cdot\text{Days}^{-1}$ to 2.5 $\text{km}^2\cdot\text{Days}^{-1}$}, there is a significant quantitative impact. In particular, for $\mu_i=.02$ and $C_0=50$, increasing $\nu_i$ increases transmission in Bergamo, Brescia and Milan by 18\%, 17\%, and 21\%, respectively. In contrast, transmssion decreases in Cremona by 10\% and in Lodi by 7\%.}

\par Finally, the capacity parameter $C_0$ \textit{did not appear to have a significant qualitative effect, with the overall trends changing little as $C_0$ is increased for specific combinations of $\nu_i$ and $\mu_i$. However, there is some quantitative impact. When $\nu_i=1.0$ and $\mu_i=.01$, increasing $C_0$ from 50 to 200 decreased transmissions by 14\% in Bergamo and 10\% in Lodi, had little effect in Brescia and Cremona, and increased transmissions by 10\% in Milan. These effects were similar across different diffusion and chemotaxis parameters. }
\par
\begin{figure}[ht!]
    \centering
    \includegraphics[width=\textwidth]{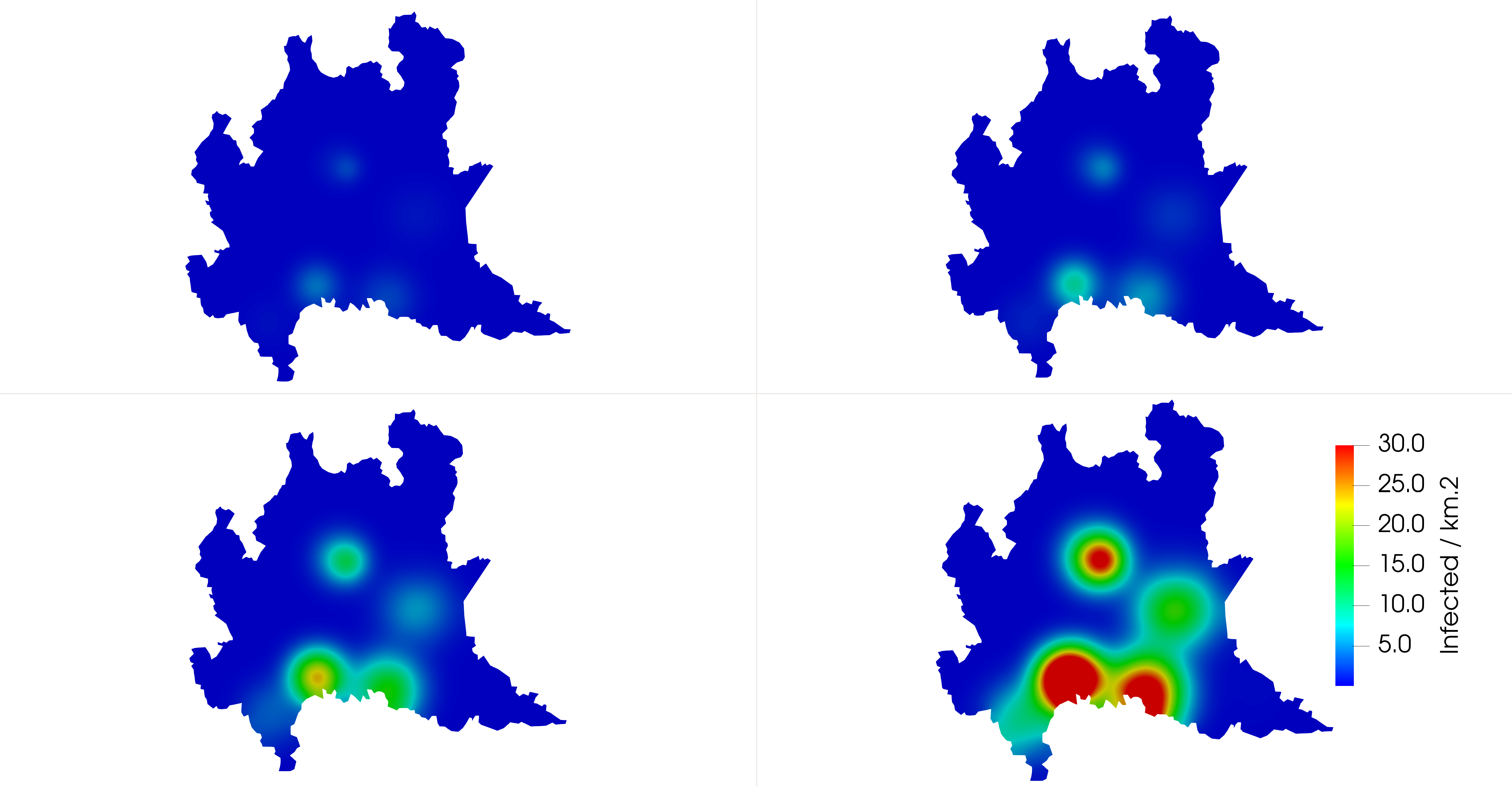}
    \caption{Left-to-right, top-to-bottom: The infected compartment on days 1, 5, 10, and 15 for $\nu_i=1.0$, $\mu_i=0$  }
    \label{fig:first15days_chem0}
\end{figure}

\begin{figure}[ht!]
    \centering
    \includegraphics[width=\textwidth]{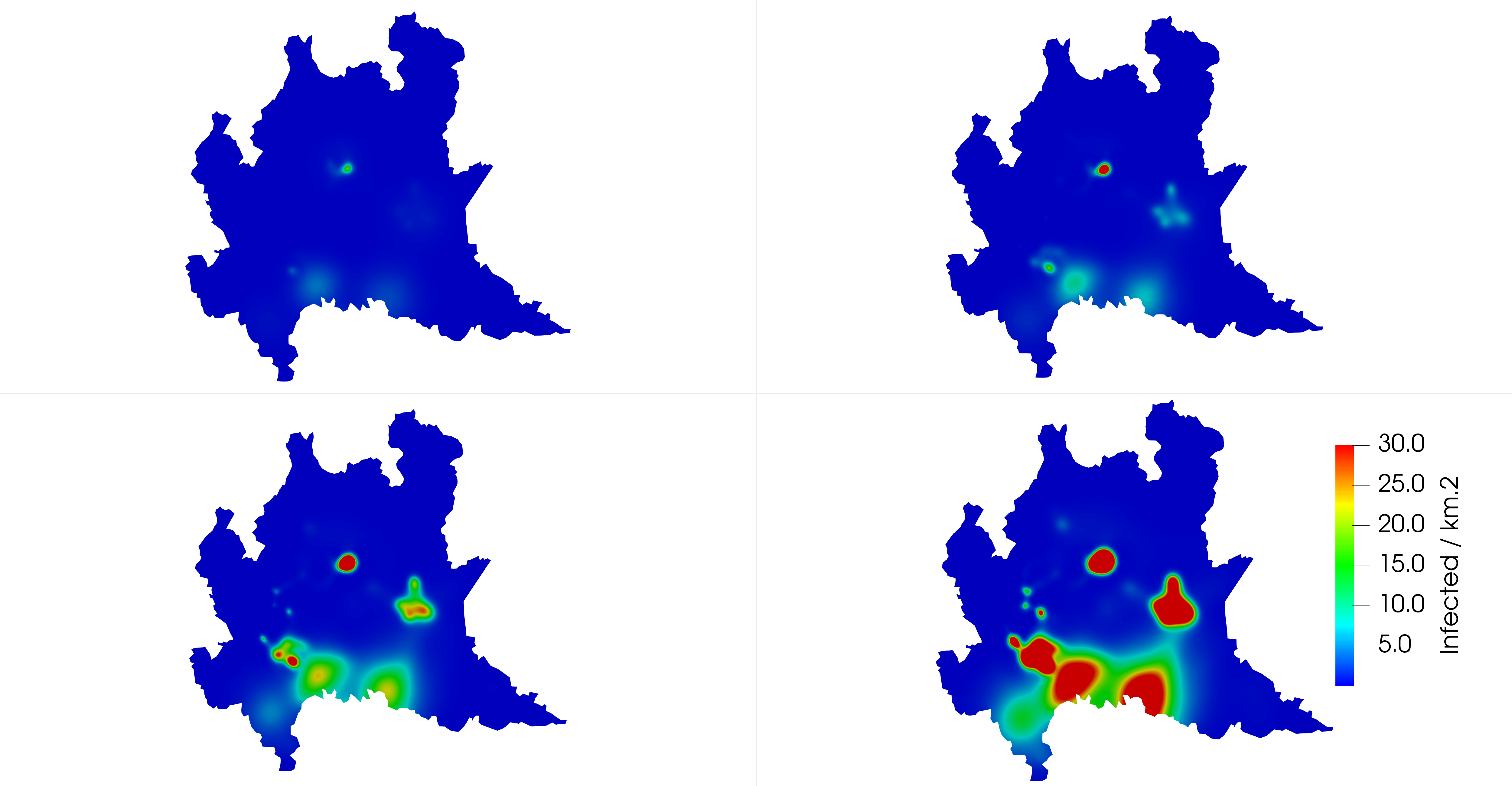}
    \caption{Left-to-right, top-to-bottom: The infected compartment on days 1, 5, 10, and 15 for $\nu_i=1.0$, $\mu_i=0.01$  }
    \label{fig:first15days_chem1}
\end{figure}

\begin{figure}[ht!]
    \centering
    \includegraphics[width=\textwidth]{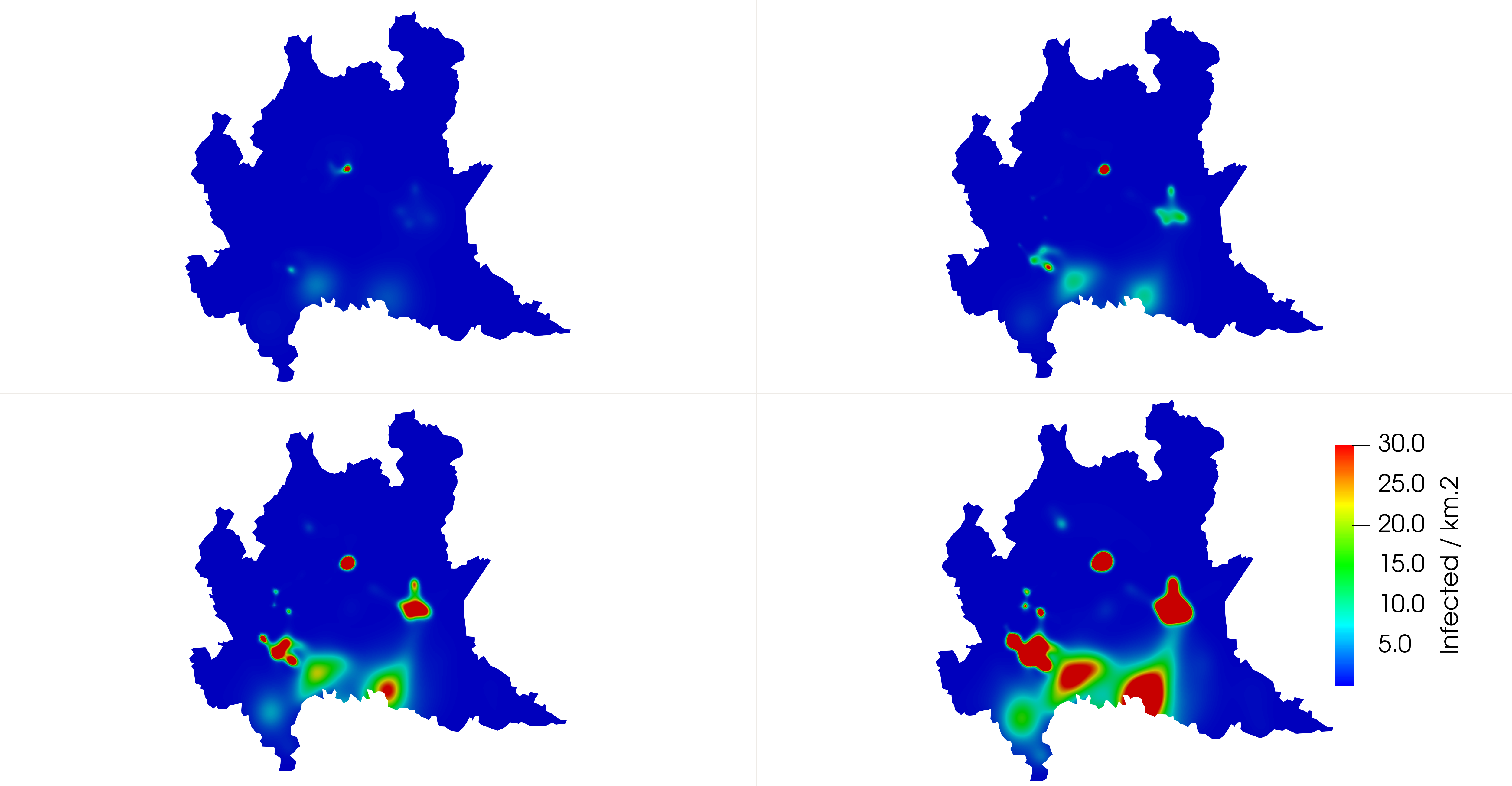}
    \caption{Left-to-right, top-to-bottom: The infected compartment on days 1, 5, 10, and 15 for $\nu_i=1.0$, $\mu_i=0.025$  }
    \label{fig:first15days_chem2}
\end{figure}

 Qualitatively, we also depict the infected compartment over the entire region for days 1, 5, 10, and 15 for $\nu_i=1.0$ and $\mu_i=0.00,\,0.01,\,0.02$ in Figs. \ref{fig:first15days_chem0}-\ref{fig:first15days_chem2}, respectively. When looking at the purely-diffusive model 
(Fig.~\ref{fig:first15days_chem0}), we see that the infection concentration remains primarily in the initially affected regions and diffuses in a regular, local, pattern, independent of the local population density. In contrast, the chemotaxis model (Figs.~\ref{fig:first15days_chem1}, \ref{fig:first15days_chem2}) shows a significantly more complex spatioteporal evolution. In particular, by day 5, we already see the appearance of nonlocalized dynamics, with the appearance of isolated hotspots in Milan. These continue to grow, and by day 15 have become a signifcant source of new transmission. Furthermore, additional pockets of transmission, appear northwest of Bergamo and Brescia (occurring in significant population centers), by day 15. In contrast, the purely diffusive model is unable to account for such dynamics, whose existence is well-supported by the available surveillance data as well as other sources (see e.g. \cite{sy2021population}).
\par Overall, our considered simulations suggest that a chemotaxis model may provide a superior description of airborne infectious disease transmission in human populations, as compared to a purely diffusive model. Our numerical simulations demonstrate that the chemotaxis model provides results more consistent with the spatiotemporal surveillance data observed in airborne epidemics. In particular, the chemotaxis model is able to better recreate the propagation of an airborne disease over large geographic distances to reach major population centers. Furthermore, these phenomena all occur as a result of the models' natural behavior, and reproducing them does not require space and/or time-dependent model parameters.

However, in practice, we note that the chemotaxis term presents significant complexity. In particular, its effect is not monotonic. Depending on the specific region, and the value of the diffusion parameter, we find that increased chemotaxis may serve to either increase or decrease transmission. This means that unlike other model parameters (for example, the contact rate), the effect of chemotaxis may present challenges when planning lockdowns/nonpharmaceutical interventions. Due to the increased complexity, a more sophisticated optimal-control approach may be necessary for such applications.

\section{Conclusions}
In the present work, we have introduced a novel PDE model for the airborne transmission of infectious disease in human populations. The novelty of the model is in the addition of a nonlinear, chemotaxis-like term to a standard reaction-diffusion model. While the standard reaction-diffusion models state that the infection should propagate from areas of high-to-low concentration of infection, the chemotaxis term further postulates that additional transmission should occur along the inverse gradient of the susceptible population; that is, from areas of low-to-high concentration of suceptible individuals. We provided an intuitive explanation motivating the reasoning behind model and literature supporting this idea. 
\par We then proceeded \betti{demonstrating} the mathematical well-posedness of the model, and show that the model provides unique solutions under 
\pier{reasonable} assumptions, for sufficiently smooth initial conditions. Furthermore, we also \betti{proved} that \pier{solutions still exist} for epidemiologically relevant parameters, including the scenario in which diffusion and chemotaxis are only considered in the infected population, and the susceptible and removed compartments are purely reactive.
\par This particular scenario was then used as the basis for a numerical simulation study, in which we simulated the propagation of an infectious disease in the Italian region of Lombardy for varying parameter values. Qualitatively, we found that the chemotaxis model was able to account for important dynamics observed in the surveillance data; however, the purely-diffusive model was unable to account for these dynamics. The results of these simulations suggest that the chemotaxis model may be more well-suited for modeling airborne infectious disease in human populations, as compared to a diffusion model. {The simulations also showed that the effect of chemotaxis on transmission is not straightforward. Depending on other factors, such as the local population density and diffusion parameters, increasing chemotaxis may either increase or reduce transmission.}
\par This work can be extended in several ways. In order to simulate a particular infectious disease at a \revis{quantitative} level, rather than just evaluating basic qualitative agreement, likely requires a more sophisticated compartmental structure, able to account for hospitalizations, \revis{asymptomatic} individuals, population demographic characteristics, and other important considerations. Nonetheless, we do not expect that these extensions will affect the \pier{mathematical results} shown herein, which should easily generalize to more complex compartmental structures. Furthermore, applying the chemotaxis model to a wider range of diseases and geographic areas is necessary to provide a further examination of the models' behavior. {Further analysis is also necessary to better understand the complex, non-monotonic relationship between chemotaxis and transmission. From a practical point of view, such understanding is necessary in order to properly apply these models in the planning and evaluation of lockdowns and nonpharmaceutical interventions. To this end, clear identification of the conditions in which chemotaxis increases/decreases transmission may be of significant practical interest, and work in this direction should be pursued.} From the numerical point of view, a more detailed analysis of numerical methods, including splitting and stabilization schemes, may be of \pier{importance} when extending the model to larger-scale problems. Finally a mathematical analysis based on \pier{some} epidemiological considerations, including the derivation of a basic reproduction number, is also \pier{a direction} for future work.

\section*{Acknowledgments}

This research activity has been performed in the framework of the
Italian-Romanian collaboration agreement ``Analysis and control of
mathematical models for the evolution of epidemics, tumors and phase field
processes'' between the Italian CNR and the Romanian Academy. In addition,
P.C. and E.R. gratefully mention some other support from the MIUR-PRIN Grant
2020F3NCPX \textquotedblleft Mathematics for industry 4.0
(Math4I4)\textquotedblright\ , the Next Generation EU Project No.P2022Z7ZAJ (A unitary mathematical framework for modelling muscular dystrophies), and the GNAMPA (Gruppo Nazionale per l'Analisi
Matematica, la Probabilit\`{a} e le loro Applicazioni) of INdAM (Istituto
Nazionale di Alta Matematica). G.M. acknowledges the support of a grant of
the Ministry of Research, Innovation and Digitization, CNCS - UEFISCDI,
project number PN-III-P4-PCE-2021-0006, within PNCD IIII.


\medskip

\begin{thebibliography}{99}

{\small

\pier{\bibitem{agnelli2023spatial}
J.~P.~Agnelli, B.~Buffa, D.~Knopoff, and G.~Torres.
\newblock A spatial kinetic model of crowd evacuation dynamics with infectious disease contagion.
\newblock {\em Bulletin of Mathematical Biology}, 85(4):23, 2023.}

\pier{\bibitem{albi2022kinetic}
G.~Albi, G.~Bertaglia, W.~Boscheri, G.~Dimarco, L.~Pareschi, G.~Toscani, and
  M.~Zanella.
\newblock Kinetic modelling of epidemic dynamics: 
social contacts, control with
  uncertain data, and multiscale spatial dynamics.
\newblock In {\em Predicting Pandemics in a Globally Connected World, Volume 1:
  Toward a Multiscale, Multidisciplinary Framework through Modeling and
  Simulation}, pages 43--108. Springer, 2022.}

\pier{\bibitem{auricchio2022well}
F.~Auricchio, P.~Colli, G.~Gilardi, A.~Reali, and E.~Rocca.
\newblock Well-posedness for a diffusion-reaction compartmental model
  simulating the spread of {COVID}-19.
\newblock {\em Mathematical Methods in the Applied Sciences},
  46(12):12529--12548, 2023.}

\pier{\bibitem{backhausCovidData} A.~Backhaus.
\newblock
Common pitfalls in the interpretation of COVID-19 data and statistics.
\newblock
{\em Intereconomics}, 55(3):162--166, 2020.}

\pier{\bibitem{bartoszek2020official} 
K.~Bartoszek, E.~Guidotti, S.~M. Iacus, and M.~Okr{\'o}j. 
\newblock Are official confirmed cases and fatalities counts good enough to study the COVID-19 pandemic dynamics? A critical assessment through the case of Italy.
\newblock {\em Nonlinear Dynamics}, 101(3):1951--1979, 2020.}

\pier{\bibitem{belik2011natural}
V.~Belik, T.~Geisel, and D.~Brockmann.
\newblock Natural human mobility patterns and spatial spread of infectious
  diseases.
\newblock {\em Physical Review X}, 1(1):011001, 2011.}

\pier{\bibitem{bellomo2022chemotaxis}
N.~Bellomo, N.~Outada, J.~Soler, Y.~Tao, and M.~Winkler.
\newblock Chemotaxis and cross-diffusion models in complex environments: Models
  and analytic problems toward a multiscale vision.
\newblock {\em Mathematical Models and Methods in Applied Sciences},
  32(4):713--792, 2022.}
 
\last{\bibitem{bertaglia2024newtrends}
G.~Bertaglia,  A.~Bondesan,  D.~Burini,  R.~Eftimie,  L.~Pareschi, and G.~Toscani. 
\newblock New trends on the systems approach to modeling SARS-CoV-2 pandemics in a globally connected planet.
\newblock {\em Mathematical Models and Methods in Applied Sciences},
 34(11):1995--2054, 2024.}
 
\pier{\bibitem{bertaglia2021hyperbolic}
G.~Bertaglia and L.~Pareschi.
\newblock Hyperbolic models for the spread of epidemics on networks: kinetic
  description and numerical methods.
\newblock {\em ESAIM: Mathematical Modelling and Numerical Analysis},
  55(2):381--407, 2021.}
  
\bibitem{blackwood2018introduction}
J.~C. Blackwood and L.~M. Childs.
\newblock An introduction to compartmental modeling for the budding infectious
  disease modeler.
\newblock {\em Letters in Biomathematics}, 5(1):195--221, 2018.

\bibitem{breda2012formulation}
D.~Breda, O.~Diekmann, W.~De~Graaf, A.~Pugliese, and R.~Vermiglio.
\newblock On the formulation of epidemic models (an appraisal of
              {K}ermack and {M}c{K}endrick).
\newblock {\em Journal of Biological Dynamics}, 6(suppl. 2):103--117, 2012.
		
\last{\bibitem{burini2024epidemics}		
D.~Burini and  D.~A. Knopoff.
\newblock Epidemics and society -- a multiscale vision from the small world to the globally interconnected world.
\newblock {\em Mathematical Models and Methods in Applied Sciences},
 34(8):1567--1596, 2024.}

\bibitem{chalub2004kinetic}
F.~A. Chalub, P.~A. Markowich, B.~Perthame, and C.~Schmeiser.
\newblock {\em Kinetic models for chemotaxis and their drift-diffusion limits}.
\newblock {\em Monatshefte f\"{u}r Mathematik},
142(1-2):123--141, 2004.

\pier{\bibitem{CGM}
P.~Colli, G.~Gilardi, and G.~Marinoschi.
\newblock Global solution and optimal control of an epidemic propagation with a
  heterogeneous diffusion.
\newblock {\em Applied Mathematics and Optimization}, 89(1):Paper No.~28, 27, 2024.}

\pier{\bibitem{CGMR-1}
P.~Colli, G.~Gilardi, G.~Marinoschi, and E.~Rocca.
\newblock Optimal control of a reaction-diffusion model related to the spread
  of {COVID}-19.
\newblock {\em Analysis and Applications}, 22(1):111--136, 2024.}

\pier{\bibitem{colombo2015smallpox}
C.~Colombo and M.~Diamanti.
\newblock The smallpox vaccine: the dispute between Bernoulli and d’Alembert and the calculus of probabilities.
\newblock {\em Lettera Matematica}, 2(4):185--192, 2015.}

\pier{\bibitem{d2024spatial}
 V.~d'Andrea, F.~Trentini, V.~Marziano, A.~Zardini, M.~Manica, G.~Guzzetta, M.~Ajelli, D.~Petrone, M.~ Del Manso, C.~Sacco, X.~Andrianou, A.~Bella, F.~Riccardo, P.~Pezzotti, P.~Poletti, and S.~Merler.
\newblock Spatial spread of COVID-19 during the early pandemic phase in Italy.
\newblock{\em BMC Infectious Diseases}, 24(1):450, 2024.}

\pier{\bibitem{D'Onofrio-al}
A.~d’Onofrio, M.~Iannelli, P.~Manfredi, and G.~Marinoschi.
\newblock Optimal epidemic control by social distancing and vaccination of an
  infection structured by time since infection: the covid-19 case study.
\newblock {\em SIAM Journal on Applied Mathematics}, 84(3):S199--S224, 2024.}

\pier{\bibitem{findlater2018human}
A.~Findlater and I.~I. Bogoch.
\newblock Human mobility and the global spread of infectious diseases: a focus
  on air travel.
\newblock {\em Trends in Parasitology}, 34(9):772--783, 2018.}

\pier{\bibitem{fitz-21}
W.~E. Fitzgibbon, J.~J. Morgan, B.~Q. Tang, and H.-M. Yin.
\newblock Reaction-diffusion-advection systems with discontinuous diffusion and
  mass control.
\newblock {\em SIAM Journal on Mathematical Analysis}, 53(6):6771--6803, 2021.}

\pier{\bibitem{fitz-18}
W.~E. Fitzgibbon, J.~J. Morgan, G.~F. Webb, and Y.~Wu.
\newblock A vector-host epidemic model with spatial structure and age of
  infection.
\newblock {\em Nonlinear Analysis. Real World Applications. An International
  Multidisciplinary Journal}, 41:692--705, 2018.}

\pier{\bibitem{gajewski1998global}
H.~Gajewski, K.~Zacharias, and K.~Gr{\"o}ger.
\newblock Global behaviour of a reaction-diffusion system modelling chemotaxis.
\newblock {\em Mathematische Nachrichten}, 195(1):77--114, 1998.}

\pier{\bibitem{gardiner2009Stochastic}
C.~Gardiner
\newblock {\em Stochastic methods}.
\newblock Springer-Verlag, Berlin, fourth edition, 2009.}

\pier{\bibitem{germann2006mitigation}
T.~Germann, K.~Kadau, I.~Longini, and C.~Macken.
\newblock Mitigation strategies for pandemic influenza in the United States.
\newblock {\em Proceedings of the National Academy of Sciences}, 103(15):5935-5940, 2006.}

\pier{\bibitem{grave2022modeling}
M.~Grave, A.~Viguerie, G.~F. Barros, A.~Reali, R.~F. Andrade, and A.~L. Coutinho.
\newblock Modeling nonlocal behavior in epidemics via a reaction--diffusion
  system incorporating population movement along a network.
\newblock {\em Computer Methods in Applied Mechanics and Engineering},
  401:115541, 2022.}

\pier{\bibitem{grave2021assessing}
M.~Grave, A.~Viguerie, G.~F. Barros, A.~Reali, and A.~L. G.~A. Coutinho.
\newblock Assessing the spatio-temporal spread of {COVID}-19 via compartmental
  models with diffusion in {I}taly, {USA}, and {B}razil.
\newblock {\em Archives of Computational Methods in Engineering. State of the
  Art Reviews}, 28(6):4205--4223, 2021.}

\pier{\bibitem{hu2013scaling}
H.~Hu, K.~Nigmatulina, and P.~Eckhoff.
\newblock The scaling of contact rates with population density for the
  infectious disease models.
\newblock {\em Mathematical Biosciences}, 244(2):125--134, 2013.}

\pier{
\bibitem{Lab24} Il Sole 24 Ore, Coronavirus in Italy: updated map and case count. 
https://lab24.ilsole24ore.com/coronavirus/en. Accessed March 10, 2024.}

\pier{\bibitem{keller2013numerical}
J.~P. Keller, L.~Gerardo-Giorda, and A.~Veneziani.
\newblock Numerical simulation of a susceptible-exposed-infectious
  space-continuous model for the spread of rabies in raccoons across a
  realistic landscape.
\newblock {\em Journal of Biological Dynamics}, 7(suppl.~1):31--46, 2013.}

\pier{\bibitem{kermack1927contribution}
W.~O. Kermack and A.~G. McKendrick.
\newblock A contribution to the mathematical theory of epidemics.
\newblock {\em Proceedings of the Royal Society of London. Series A, Containing
  papers of a mathematical and physical character}, 115(772):700--721, 1927.}

\pier{\bibitem{kim2021kinetic}
D. Kim and A. Quaini.
\newblock A kinetic theory approach to model crowd dynamics with disease contagion.
\newblock {\em Crowd Dynamics, Volume 3: Modeling and Social Applications in the Time of COVID-19}.
\newblock Springer-Verlag, 2021.}

\pier{\bibitem{li2009modeling}
J.~Li and X.~Zou.
\newblock Modeling spatial spread of infectious diseases with a fixed latent
  period in a spatially continuous domain.
\newblock {\em Bulletin of Mathematical Biology}, 71:2048--2079, 2009.}

\pier{\bibitem{Lions}
J.-L. Lions.
\newblock {\em \'{E}quations diff\'{e}rentielles op\'{e}rationnelles et
  probl\`emes aux limites}.
\newblock Die Grundlehren der mathematischen Wissenschaften, Band 111.
  Springer-Verlag, Berlin-G\"{o}ttingen-Heidelberg, 1961.}

\pier{\bibitem{Lions-69}
J.-L. Lions.
\newblock {\em Quelques m\'{e}thodes de r\'{e}solution des probl\`emes aux
  limites non lin\'{e}aires}.
\newblock Dunod, Paris; Gauthier-Villars, Paris, 1969.}


\pier{\bibitem{marinoschi2013well}
G.~Marinoschi.
\newblock Well-posedness for chemotaxis dynamics with nonlinear cell diffusion.
\newblock {\em Journal of Mathematical Analysis and Applications},
  402(2):415--439, 2013.}

\pier{\bibitem{GM-AMO}
G.~Marinoschi.
\newblock Parameter estimation of an epidemic model with state constraints.
\newblock {\em Applied Mathematics and Optimization}, 84(suppl.
  2):S1903--S1923, 2021.}

\pier{\bibitem{GM-DCDS}
G.~Marinoschi.
\newblock Identification of transmission rates and reproduction number in a
  {SARS}-{C}o{V}-2 epidemic model.
\newblock {\em Discrete and Continuous Dynamical Systems. Series S},
  15(12):3735--3744, 2022.}
  
\pier{%
\bibitem{GM-NA}
G.~Marinoschi.
\newblock A semigroup approach to a reaction-diffusion system with
  cross-diffusion.
\newblock {\em Nonlinear Analysis. Theory, Methods \& Applications. An
  International Multidisciplinary Journal}, 230:Paper No. 113222, 29, 2023.}

\pier{\bibitem{murray2003mathematical}
J.~D. Murray.
\newblock {\em Mathematical biology. {II}}, volume~18 of {\em Interdisciplinary
  Applied Mathematics}.
\newblock Springer-Verlag, New York, third edition, 2003.
\newblock Spatial models and biomedical applications.}

\pier{\bibitem{negreanu2014two}
M.~Negreanu and J.~I. Tello.
\newblock On a two species chemotaxis model with slow chemical diffusion.
\newblock {\em SIAM Journal on Mathematical Analysis}, 46(6):3761--3781, 2014.}

\pier{\bibitem{piazzola-21}
C.~Piazzola, L.~Tamellini, and R.~Tempone.
\newblock A note on tools for prediction under uncertainty and identifiability of SIR-like dynamical systems for epidemiology.
\newblock {\em Mathematical Biosciences}, 332:108514, 2021.}

\pier{\bibitem{ramaswamy2021comprehensive}
H.~Ramaswamy, A.~A. Oberai, and Y.~C. Yortsos.
\newblock A comprehensive spatial-temporal infection model.
\newblock {\em Chemical Engineering Science}, 233:116347, 2021.}

\pier{\bibitem{SSalsa}
S.~Salsa.
\newblock {\em Partial Differential Equation in Action}, 4th Edition. 
\newblock Springer, Milan, 2015.}

\pier{\bibitem{schiesser2018mathematical}
W.~E. Schiesser.
\newblock {\em Mathematical Modeling Approach To Infectious Diseases, A: Cross
  Diffusion {PDE} Models For Epidemiology}.
\newblock World Scientific, 2018.}

\pier{\bibitem{s2012radiation}
F.~Simini, M.~Gonzalez, A.~Maritan, and A.~Barabasi.
\newblock A universal model for mobility and migration patterns.
\newblock{\em Nature}, 484:96-100, 2012.}

\pier{\bibitem{sy2021population}
K.~T.~L. Sy, L.~F. White, and B.~E. Nichols.
\newblock Population density and basic reproductive number of covid-19 across
  united states counties.
\newblock {\em PloS one}, 16(4):e0249271, 2021.}

\pier{\bibitem{tatapudi2022evaluating}
H.~Tatapudi and C.~Gopalappa.
\newblock Evaluating the sensitivity of jurisdictional heterogeneity and
  jurisdictional mixing in national level hiv prevention analyses: context of
  the US Ending the HIV Epidemic plan.
\newblock {\em BMC Medical Research Methodology}, 22(1):304, 2022.}

\pier{\bibitem{viguerie2022coupled}
A.~Viguerie, G.~F. Barros, M.~Grave, A.~Reali, and A.~L. Coutinho.
\newblock Coupled and uncoupled dynamic mode decomposition in
  multi-compartmental systems with applications to epidemiological and additive
  manufacturing problems.
\newblock {\em Computer Methods in Applied Mechanics and Engineering},
  391:114600, 2022.}

\pier{\bibitem{viguerieNumSim} A.~Viguerie, A.~Veneziani, G.~Lorenzo, D.~Baroli, N.~Aretz-Nellesen, A.~Patton,
  T.~E. Yankeelov, A.~Reali, T.~J. Hughes, and F.~Auricchio.
\newblock Diffusion--reaction compartmental models formulated in a continuum
  mechanics framework: application to covid-19, mathematical analysis, and
  numerical study.
\newblock {\em Computational Mechanics}, 66:1131--1152, 2020.}

\pier{\bibitem{zanella2024derivation}
M.~Zanella.
\newblock Derivation of macroscopic epidemic models from multi-agent systems.
\newblock Preprint arXiv:2410.08610 [q-bio.PE], 2024.}

\pier{\bibitem{zhao2023spatiotemporal}
G.~Zhao and S.~Ruan.
\newblock Spatiotemporal dynamics in epidemic models with {L}\'{e}vy
              flights: a fractional diffusion approach.
\newblock {\em Journal de Math\'{e}matiques Pures et Appliqu\'{e}es. Neuvi\`eme S\'{e}rie},
  173:243--277, 2023.}

} 

\end{thebibliography}
\end{document}